\documentclass[twocolumn]{jpsj2} %% two-column layout
\usepackage{color}
\usepackage{multirow}
\usepackage{graphicx}  % Include figure files
\usepackage{dcolumn}   % Align table columns on decimal point
\usepackage{bm}        % bold math
\usepackage{color}
\newcommand{\prl}{\rm Phys. Rev. Lett.}

\newcommand{\prb}{\rm Phys. Rev. B}
\newcommand{\rmp}{\rm Rev. Mod. Phys.}
\def\simge{\lower0.7ex\hbox{$\ \overset{>}{\sim}\ $}}
\def\simle{\lower0.7ex\hbox{$\ \overset{<}{\sim}\ $}}

\title{Classical Monte Carlo Study for Antiferro Quadrupole Orders in a Diamond Lattice}

\author{Kazumasa Hattori and Hirokazu Tsunetsugu}

\inst{Institute for Solid State Physics, University of Tokyo, 5-1-5, Kashiwanoha, Kashiwa, Chiba 277-8581, Japan
}

\author{Kazumasa Hattori$^{1,2}$\thanks{E-mail:
hattori@tmu.ac.jp} and Hirokazu Tsunetsugu$^1$}
\inst{$^1$The Institute for Solid State Physics, The University of Tokyo, 5-1-5, Kashiwanoha, Kashiwa, Chiba 277-8581, Japan \\
$^2$Department of Physics, Tokyo Metropolitan University, 
1-1 Minami-osawa, Hachioji, Tokyo 192-0397, Japan}

\abst{We investigate antiferro quadrupole orders in a diamond lattice 
under magnetic fields by Monte Carlo simulations 
for two types of classical effective models. 
One is an XY model with $Z_3$ anisotropy, and the other is a two-component 
$\phi^4$ model with a third-order anisotropy.
We confirm that the universality class of the zero-field 
transition is that for the three-dimensional XY model. 
Magnetic field corresponds to a $Z_3$ field in the effective model, 
and under this field, 
we find that collinear and canted antiferro-quadrupole orders compete. 
Each phase is characterized by symmetry breaking 
in the sector of 
(sublattice $Z_2)\otimes$(reflection $Z_2$ for the order parameter). 
When $Z_3$ anisotropy and magnetic field vary, it turns out 
that this system is a good playground for various multicritical points; 
bicritical and tetracritical points emerge in a finite field. 
Another important finding is about the scaling of parasitic ferro quadrupole 
order at the zero-field critical point.  
This is the secondary order parameter induced by the primary antiferro 
order, and its critical exponent $\beta'$=0.815 clearly differs from 
the expected value that is twice the value for the primary order parameter.  
The corresponding correlation length exponent is also different, 
$\nu'$=0.597(12).  
We also discuss relation of the present effective quadrupole models 
with the 3-state Potts model as well as implication to 
undertanding of orbital orders in Pr-based 1-2-20 compounds.  
 }

\kword{orbital order, quadrupole, Monte Carlo simulation, Potts model}

\begin{document}
\maketitle

%%%%%%%%%%%%%%%%%%%%% Introduction %%%%%%%%%%%%%%%%%%%%%%%%%%%%%%%%%%%%%%%%%%%%
\section{Introduction}
Orbital physics in strongly correlated electron systems has been 
intensively studied in recent years \cite{Review,Kuramoto}. 
The orbital degrees of freedom in partially-filled \textit{d}- or 
\textit{f}-electron levels show a variety of interesting 
phases and properties. 
Anisotropic nature of these electron wavefunctions with nonzero
angular momentum is a key to understanding these materials, 
and this leads to, for example, a strongly anisotropy in the spin space in
addition to the orbital sector itself \cite{Kugel, Kitaev, Jackeli}.

Recently, orbital degrees of freedom in Pr-based $f$-electron
systems Pr$T$$_2$$X_{20}$ ($T$=V, Ti, Ir, Rh and $X$=Al, Zn) have attracted 
great attention.   
Quadrupolar Kondo effects \cite{Cox} are expected to 
take place at Pr ions with non-Kramers-doublet ground state 
\cite{Onimaru0,Onimaru,Onimaru2,Sakai1,Sakai2}. 
Electric resistivity approaches the zero-temperature limit 
with a negative finite-temperature correction that 
is singular at $T$=0. 
This is naively attributed to a local quadrupole Kondo 
effect \cite{Sakai1,Sakai2}, and 
there have been a few theoretical developments 
about the two-channel Kondo lattice 
systems \cite{Tsuruta,Hoshino}. 
A microscopic model is also proposed for describing 
these compounds \cite{Kusu}.

Those compounds also have one or a few low-temperature phases 
at zero magnetic field and also multiple phases under magnetic fields. 
Most of them are considered to be antiferro quadrupole ordered phases.  
An exception is PrTi$_2$Al$_{20}$\cite{Sakai1,Sakai2}, and 
the neutron scattering \cite{Sato} and ultrasound 
experiments \cite{Koseki} suggest a ferro quadrupole order in this compound. 
Anisotropy in the orbital sector in these systems is manifested 
in the strong anisotropy of the critical field 
and the phase diagram strongly dependent on the field direction 
\cite{Onimaru2, Sakai1,Ishii1,Ishii2}.
Superconductivity appears in a couple of compounds near the quadrupole 
ordered phases \cite{Onimaru0,Matsubayashi}, and it is expected that 
orbital fluctuations in the non-Kramers
doublet in the Pr ions contribute to its realization.

In our previous study~\cite{Hattori}, antiferro orbital orders 
in the Pr 1-2-20 compounds were investigated.  
Using a mean-field and spin-wave theories, we determined 
the temperature-magnetic field phase diagram, and 
also calculated excitation spectra.   
Quantum effects at each Pr ion were taken into account 
by considering all the crystalline-electric-field (CEF) states 
within the $J=4$ multiplet in the calculations, 
while the intersite correlations were approximated 
by static mean fields.  

The purpose of the present work
is to examine the effects of intersite thermal fluctuations 
on the antiferro quadrupole orders and the phase transitions, and 
we employ classical Monte Carlo (MC) method to this end. 
Since the transition temperatures of the ordered phases 
are much lower than the energy scales of the 
CEF excited states \cite{Iwasa}, it is justified to use 
an effective classical model constructed in terms of the
non-Kramers doublet CEF ground states alone. 
In addition to antiferro quadrupole orders, 
ferro quadrupole and octupole orders are other possibilities 
of symmetry breaking, but we will not examine them in this paper. 
We should note that antiferro orders exhibit much richer physics 
than ferro orders especially in magnetic field.  
A challenging subject of octupole orders requires a formulation 
more complicated than those outlined in this paper, and 
therefore we leave this a future study.  

We will introduce in Sec.~\ref{sec-Model} a system to be investigated 
and that is a model of the $E$-modes of electric quadrupoles in the Pr compounds.  
We will show that 
its classical effective model is a plane rotor (or XY) 
model with a $Z_3$ single-site anisotropy, or if amplitude fluctuations are 
not traced out, a $\phi^4$ model for a two-component field 
with a third-order term.  
Having the same symmetry, these two effective models are related to 
the 3-state Potts model (equivalent to the 3-state clock model), 
and this simpler model has been intensively studied in the community 
of statistical physics for the case of ferro interactions\cite{Potts}. 
In contrast, the low-temperature ordered phases 
are not well understood when interactions are antiferro. 
There are controversies in their properties among the reported studies 
as we will explain later.  

Another purpose of this paper is to achieve better understanding of 
the ordered phases of the antiferro 3-state Potts model on the basis of 
 our calculations.  
Several previous works have studied  
the antiferro 3-state Potts model on three-dimensional bipartite 
lattices,\cite{Banavar,Grest,Rosengren,Kolesik,Pelizzola,Heilmann,Ni,Lapinskas,Rahman,Okabe} 
and their results have agreed that the anisotropy is 
irrelevant at the phase transition and the transition 
belongs to the XY universality class in three 
dimensions\cite{Banavar}. 
However, their results are not consistent to each other 
about the ordered phases, and the nature of the ordered states 
has not been well clarified \cite{Banavar,Grest,Rosengren,Kolesik,Pelizzola,Heilmann,Ni,Lapinskas,Rahman,Okabe}.   
In our models, microscopic degrees of freedom of local order parameters 
are enlarged from three points in the Potts model, 
and can be located a unit circle (one-dimensional compact manifold) 
or a two-dimensional continuous vector space 
(two-dimensional noncompact manifold).  
This corresponds to coarse graining processes of the Potts model, and 
we expect that our models exhibit the nature of order parameter 
configuration more clearly.

This paper is organized as follows. In Sec.~\ref{sec-Model}, a classical
quadrupole Hamiltonian in a diamond lattice is introduced. We have 
analyzed the model by classical Monte Carlo simulations and the numerical
results for zero magnetic field will be shown in
Sec.~\ref{sec-zeromag}. Detailed analysis of ordered states will be
carried out in Sec.~\ref{sec-LT} with comparison to the antiferro
3-state Potts model. Effects of magnetic field will be examined in
Sec.~\ref{sec-THphase} and the temperature-magnetic field phase diagram
will be constructed. 
The relevance of the
present results on the Pr-based 1-2-20 systems are discussed in
Sec.~\ref{sec-Discussion}, and finally Sec.~\ref{sec-Summary} concludes the
present paper.

\section{Models and Numerical Method}
\label{sec-Model}

In this section, we briefly explain the characteristics of the
non-Kramers doublet system in the Pr 1-2-20 compounds 
and introduce two model Hamiltonians to be studied in this paper. 

\subsection{Non-Kramers doublet for Pr ion}
In the Pr 1-2-20 compounds, the Pr ions form a diamond sublattice.  
Each ion has valency Pr$^{3+}$ and two
f-electrons. On the basis of the LS coupling scheme, this electron configuration 
has the total angular momentum $J=5-1=4$. The local point group symmetry
at the Pr site is $T_d$, and the CEF potential splits  
the $J=4$ multiplet.  
It is known that the ground state is a non-Kramers doublet with $E$($\Gamma_3$)
representation 
\cite{Iwasa}. 
Its basis states are 
\begin{eqnarray}
|\Uparrow\rangle&=&\frac{1}{\sqrt{12}}\Big[
\sqrt{7}\, \frac{|4\rangle+|-4\rangle}{\sqrt{2}}-\sqrt{5}|0\rangle\Big],\\
|\Downarrow\rangle &=& -\frac{|2\rangle+|-2\rangle}{\sqrt{2}},
\label{eq2:basis}
\end{eqnarray}
where $|J_z\rangle$ denotes the eigenstate of the total angular momentum $J=4$ and its 
$z$-component $J_z$. 
Note that these two states are both invariant upon 
time-reversal operation, and thus, the ground state is a non-Kramers doublet.

The internal dynamics in the non-Kramers doublet is completely described 
by three operators usually 
denoted by the three Pauli matrices $(\sigma_x,\sigma_y,\sigma_z)$.   
Two of them are quadrupole operators 
$\sigma_z = \mathcal{P}_0 \frac{1}{8} ( 2J_z^2 - J_x^2 - J_y^2 ) \mathcal{P}_0$ 
and 
$\sigma_x =  \mathcal{P}_0 \frac{\sqrt{3}}{8} ( J_x^2 - J_y^2 ) \mathcal{P}_0$,  
where $\mathcal{P}_0$ is the projector onto the non-Kramers doublet 
ground state.  
They operate two basis states in the ground-state doublet as 
\begin{eqnarray}
&&\sigma_x |\Uparrow\rangle = |\Downarrow\rangle , \ \ 
\sigma_x |\Downarrow\rangle = |\Uparrow\rangle , 
\nonumber\\
&&\sigma_z |\Uparrow\rangle = |\Uparrow\rangle , \ \ 
\sigma_z |\Downarrow\rangle = -|\Downarrow\rangle . \ \ 
\end{eqnarray} 
The last one $\sigma_y$ is an octupole
moment $\sigma_y =   \mathcal{P}_0 \frac{-1}{36\sqrt{3}}  [J_x J_y J_z + \mbox{(all permutations)}]  \mathcal{P}_0 $. As mentioned in
the Introduction, we will concentrate on the quadrupole degrees of freedom $(\sigma_z, \sigma_x)$
alone in this paper.

\subsection{Classical mapping and single-ion anisotropy}
For  finite-temperature phase transitions, the most of essential 
fluctuations arise from thermal ones. 
In our previous study,\cite{Hattori} 
we have analyzed a microscopic model for the Pr 1-2-20 compounds 
by means of the mean-field theory and the spin-wave type analysis. 
The scheme based on the microscopic model takes into account the 
full information of the CEF states for the Pr ions, 
while it is insufficient for determining the universality class 
of the transition, since intersite fluctuations are not 
taken into account.  
In order to determine a phase diagram and universality class of the transitions, 
we will construct effective 
classical models and analyze them with fully taking account of 
thermal fluctuations.  
In the following, we will introduce two effective models 
and study them throughout this paper

The classical correspondence of the pair of quadrupole operators 
$(\sigma_z,\sigma_x)$ at each Pr site is a two-component classical vector 
$\mathbf{Q}=(Q_u,Q_v)$.  
With symmetry operations of the $T_d$ point group, 
its two elements transform as bases of the $E$ representation, and 
they have the same symmetry as 
$Q_u\sim 2z^2-x^2-y^2$ and $Q_v\sim \sqrt{3}(x^2-y^2)$.  
We also use the polar representation $\mathbf{Q}=Q(\cos\theta,\sin\theta)$. 

For considering effective interactions between quadrupole moments, 
it is important to count all the possible low-order invariants made of them.  
Local invariants are straightforwardly
obtained, and some of them describe local anisotropy. Apart from the trivial 
quadratic term $\propto |\mathbf{Q}|^2$, it should be
noted that the single ion terms contain a third-order anisotropy 
[see eq. (\ref{H234})].
This $Z_3$ anisotropy corresponds to the three choices of the uniaxial 
direction of the quadrupole moment, namely $x$, $y$, or $z$ directions. 
It should be noted that this type of anisotropy does not exist, 
if the local order parameter is a magnetic dipole, since 
it is not invariant upon time reversal operation.  
Therefore, this anisotropy is specific to the  quadrupole order parameters.
Indeed, as discussed in Ref. \citenum{Hattori}, the magnitude of 
this anisotropy is about 
$\sim$1 K in Pr-based 1-2-20 compounds, and this is
comparable to the transition temperature for the quadrupole
orders\cite{Onimaru0,Onimaru,Onimaru2,Sakai1,Sakai2}.

Another point we should
mention is the coupling to magnetic fields. Since the quadrupoles are 
nonmagnetic degrees of freedom, they do not couple to magnetic 
fields in the linear order.  
However, there exists a quadratic ``Zeeman'' coupling as will be
introduced in Sec. \ref{sec-THphase}, which is important for understanding
the phase diagram of the Pr 1-2-20 compounds under magnetic fields.  

\subsection{Model Hamiltonian}
In this paper, we are going to study a system of 
interacting quadrupoles 
$\{ \mathbf{Q}_i  \}$ 
on the diamond lattice.  
Here, $\mathbf{Q}_i = Q_i (\cos \theta_i , \sin \theta_i )$ 
represents two components of the 
quadrupole moment of Pr ion at the site $i$. 
In the following, we will call ${\bf Q}_i$ simply a pseudo spin, 
or otherwise explicitly a quadrupole moment.
For this system, we first consider the Hamiltonian defined as 
\begin{eqnarray}
&&H=H_{\rm loc}+H_{\rm int},
\label{Model} \\ 
&&H_{\rm loc} = \sum_i \left( \frac{a}{2}Q^2_i
+\frac{b}{4}Q^4_i
-\frac{c}{3}Q_i^3\cos3\theta_i
\right) ,
\label{H234}\\
&&H_{\rm int}
=J\sum_{\langle i,j\rangle}Q_i Q_j\cos(\theta_i-\theta_j) 
=J\sum_{\langle i,j\rangle} \mathbf{Q}_i \cdot \mathbf{Q}_j.\ \ \ \   \label{Hint}
\end{eqnarray}
Note that magnitudes of quadrupole moments are variables in this model, 
and they fluctuate.  
In the interaction part $H_{\rm int}$, the coupling is an antiferro type and 
the sum is taken over all 
the nearest-neighbor bonds.  
It is a special feature of this system that 
only this isotropic coupling is allowed. 
This is because each Pr-Pr bond points along [111] or one of 
its equivalent directions and this restricts a possible form of 
intersite quadrupole coupling\cite{Hattori}.  
Describing the symmetries of the system properly, 
this Hamiltonian is sufficient for studying quadrupole orderings, 
while higher-order terms in $Q$'s are safely neglected.  
As for the coupling constants, we set $a=-5$, $b=10$ and the unit of
energy is $J=1$, throughout this paper. 
In Secs. \ref{sec-zeromag} and \ref{sec-LT}, we will also 
study a further simplified model.  
Fluctuations of the amplitudes are traced out there 
and we set $Q_i = 1$ at all the sites, and each quadrupole moment 
has the angle degrees of freedom $\theta_i$ alone.  
Thus, the simplified Hamiltonian is 
\textit{a plane rotor model with three-fold anisotropy}:
\begin{eqnarray}
H=-\frac{c}{3}\sum_i\cos3\theta_i+J\sum_{\langle i,j\rangle}\cos(\theta_i-\theta_j). \label{eq-rotor}
\end{eqnarray}
The models (\ref{Model}) and (\ref{eq-rotor}) contain only the quadrupole 
degrees freedom of the localized $f$-electrons and we will not
discuss their coupling to conduction electrons, which is important in 
the Kondo physics observed in the experiments.\cite{Sakai1} As for the quadrupole 
ordering, however, these models capture its essential 
aspects in the Pr-based 1-2-20 compounds.

In the following sections, we will present our numerical results in the 
classical Monte Carlo (MC) simulations for the models (\ref{Model}) 
and (\ref{eq-rotor}).  
The typical system size defined 
by the number of sites is 
$N = L^3$ $(8 \le L \le 128)$, and the periodic boundary condition is used for all the three directions. To be specific, $L=2$ corresponds to the cubic
unit cell, which contains 8 sites. For updates in the MC simulation, 
we combined the Metropolis algorithm of 
single-site flips and the Wolff's algorithm\cite{Wolff}.  
We also use several global updates; the global $C_3$ rotation ($\theta_i\to \theta_i\pm 2\pi/3$) of $\mathbf{Q}$ when the applied magnetic field is zero or weak, 
or otherwise the global update of $Q_v\to -Q_v$ 
[see eq.~(\ref{Hmag}) for coupling to the magnetic field].  
In the simulations, each MC run sampled $\sim$50,000 snapshots 
after thermalization typically 50,000 MC steps, and the data were averaged over typically 64 
MC runs starting from different initial configurations.

\section{Phase Transition at Zero Magnetic Field} 
\label{sec-zeromag}

Let us start studying a phase transition in our quadrupole systems  
from the case of zero magnetic field. 
Our previous mean-field analysis \cite{Hattori} showed
that the system undergoes 
a phase transition with decreasing temperature 
into an ordered phase. There, the $Z_3$ symmetry in the 
quadrupole pseudospin space is spontaneously broken.  
The transition is continuous unless the anisotropy is too large. 
The mean-field theory also shows that 
the primary order parameter is the staggered $Q_v$ component 
and that this is accompanied by the secondary order parameter, 
the uniform $Q_u$ component. 
These order parameters are about one of the degenerate ordered phases.  
The ordered phase has degeneracy $6 =3 \times 2$, where 
3 comes from $Z_3$ symmetry and 2 comes from the symmetry between 
$A$- and $B$-sublattices.  
In the four other ordered phases, the primary and secondary 
order parameters are rotated by $\pm 2\pi /3$ in 
the $\mathbf{Q}$ space.  

We will study the effects of thermal fluctuations on 
the phase transition by using classical Monte-Carlo simulations 
and confirm the symmetry breaking predicted by the mean-field theory. 
We will next investigate critical behaviors of the phase transition.  
Several previous works have studied this problem of 
the antiferro 3-state Potts model on three-dimensional bipartite lattices. 
Their conclusion is that the anisotropy is irrelevant and 
the transition belongs to the XY universality class in three 
dimensions,\cite{Banavar} 
{\it i.e.}, that of superfluid transition in liquid ${}^4$He.  
Large-scale numerical simulations have been performed to study 
this universality class, and the values of critical exponents 
are determined in high precision:\cite{He4}  
$\eta$=$0.0381(2)$, $\nu$=$0.6717(1)$, 
$\alpha$=$-0.0151(3)$, $\gamma$=$1.3178(2)$, 
$\beta$=$0.3486(1)$, and $\delta$=$4.780(1)$.
We will also check if our MC results are consistent 
with these exponents.  

\subsection{Monte Carlo calculations}

In MC simulations, we calculated 
thermodynamic quantities including specific heat, 
order parameters, and correlation functions. 
For order parameters, the average $\mathbf{Q}_{s}$ 
is first defined for each sublattice $s$ and 
then the staggered and the uniform components are defined as
\begin{equation} 
\mathbf{Q}_{\mp} = \frac{\mathbf{Q}_{A} \mp \mathbf{Q}_{B} }{2} , \ \ 
\mathbf{Q}_{s} = \frac{2}{N} \sum_{j \in s} \mathbf{Q}_j , \ \ 
(s = A, B). 
\label{eq3:Q}
\end{equation}
With this normalization, their variance of fluctuation 
$\sqrt{ \langle \mathbf{Q}_{\mp}^2 \rangle \, }$ 
plays the role of order parameters. 
In the limit of $N \rightarrow \infty$, the order parameter 
is expected 
to vanish in the high-temperature para phase and 
stays finite in the low-temperature ordered phase.  

As for the correlation functions, we will analyze  their 
Fourier transform.  
The primitive unit cell consists of two sites, $A$ and $B$, 
and each quadrupole moment has two components 
$(Q_u ,Q_v)$.  
Therefore, the Fourier transform of the correlation function 
is a $4 \times 4$ matrix 
\begin{equation}
 C_{ s \mu , s' \mu' } (\mathbf{k}) 
 = \frac{1}{N/2} \sum_{j \in s} \sum_{j' \in s'} \langle 
Q_{\mu ,j } Q_{\mu' , j' } \rangle 
e^{-i \mathbf{k} \cdot (\mathbf{r}_{j} - \mathbf{r}_{j'})},
\end{equation}
where $s$, $s'$ are the sublattice index and 
$\mu$, $\mu'$ are the pseudospin component $u$ or $v$.  
This is a hermitian matrix and all of its four 
eigenvalues $\{ \lambda_n (\mathbf{k}) \}$ are real.  
The wave vector $\mathbf{k}^*$ where the largest eigenvalue 
$\lambda_1 (\mathbf{k})$ is maximum 
and its eigenvector describe the most dominant 
spatial correlation of quadrupoles. 
The correlation length $\xi$ is a very important quantity 
for investigating critical properties, and 
we define it by the peak width of the largest eigenvalue 
$\lambda_1 (\mathbf{k})$  as 
\begin{equation}
\frac{\lambda_1 (\mathbf{k}_1 )}{\lambda_1 (\mathbf{k}^* )} 
= 1 - 4 \xi^2 \sin^2 \frac{\delta k}{2} , 
\ \ \ 
\mathbf{k}_1 = \mathbf{k}^* +  (\delta k , 0, 0 ), 
\end{equation}
where $\mathbf{k}_1$ is the wave vector closest to the peak 
position $\mathbf{k}^*$ in the finite size system considered.  

%%%%%%%%%%%%%%%%%%%%%%%%%%% fig 1 %%%%%%%%%%%%%%%%%%%%%%%
\begin{figure}[t]
\begin{center}
\includegraphics[width=0.45\textwidth,clip]{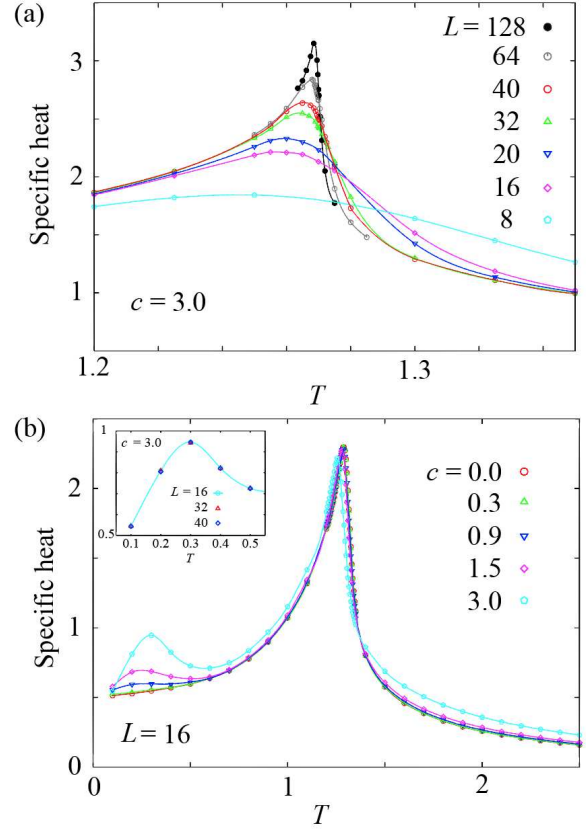}
\end{center}
\caption{ (Color online)
Temperature dependence of specific heat. 
(a) Near $T$=$T_c$ for $c$=3.0 and $8 \le L \le 128$. 
(b) $c$ dependence for $L=16$. 
Inset in (b): the system size dependence of low-temperature peak 
for $c$=3.0.}
\label{fig:CT}
\end{figure}
%%%%%%%%%%%%%%%%%%%%%%%%%%%%%%%%%%%%%%%%%%%%%%%%%%%%%%%%%

\subsection{Thermodynamics}

Figure \ref{fig:CT}(a) shows the temperature dependence 
of specific heat $C(T)$ 
in the rotor model (\ref{eq-rotor}) with the anisotropy $c=3.0$.   
Each curve of the data for the system size $8 \le L \le 128$ 
has a peak around the temperature $T=1.27$. 
As the system size increases, the peak sharpens.  
This indicates the presence of a phase transition, 
but the peak value does not grow unlike many other transitions.  
This behavior is consistent with the expectation about 
the specific heat critical exponent $\alpha < 0$.  
Another possibility is that the transition is first order, 
but we have excluded this possibility by calculating internal 
energy $E(T)$. 
It does not show a jump at the transition temperature, 
and thus, we conclude that the transition is continuous. 
We will examine the critical exponent later.

We also quickly examine the effects of anisotropy $c$.  
Figure \ref{fig:CT}(b) shows the specific heat $C(T)$ 
for several values of the anisotropy from $c=0.0$ to $3.0$ 
calculated for the system size $L=16$.   
The limit $c=0.0$ is the isotropic XY model 
with antiferro nearest-neighbor couplings 
on the diamond lattice, and its $C(T)$ should agree 
with the one of the ferro XY model,  
because of its bipartite lattice structure and 
no field applied.  
Critical behavior in the specific heat does not seem to 
depend sensitively on the value of the anisotropy 
up to $c = 3.0$.  

The specific heat $C(T)$ also has a broad peak 
at low temperature $T \sim 0.3$, and it grows as $c$ increases. 
A few previous theories predicted the presence of multiple 
phase transitions in a model related to ours \cite{Banavar,Grest,Rosengren,Kolesik,Pelizzola,Heilmann,Ni,Lapinskas,Rahman,Okabe}, but 
our result is not consistent with their prediction.
The system size dependence of this peak is shown in the inset of Fig.~\ref{fig:CT}(b) for $c=3.0$. 
Our data do not show a noticeable system size dependence 
expected at a phase transition, and thus 
this small peak is not a phase transition.
As for the origin of the peak, we have found a signature 
showing that the system gains a local anisotropy energy there.  
The temperature dependence of the 
average $\langle \cos 3\theta_i\rangle $ slightly steepens 
around the temperature at the specific heat peak.
We will discuss the comparison to the related model 
in more detail in Sec.~\ref{sec-Discussion}.

%%%%%%%%%%%%%%%%%%%%%%%%%%% fig 2 %%%%%%%%%%%%%%%%%%%%%%%
\begin{figure}[t]
\begin{center}
\includegraphics[width=0.5\textwidth]{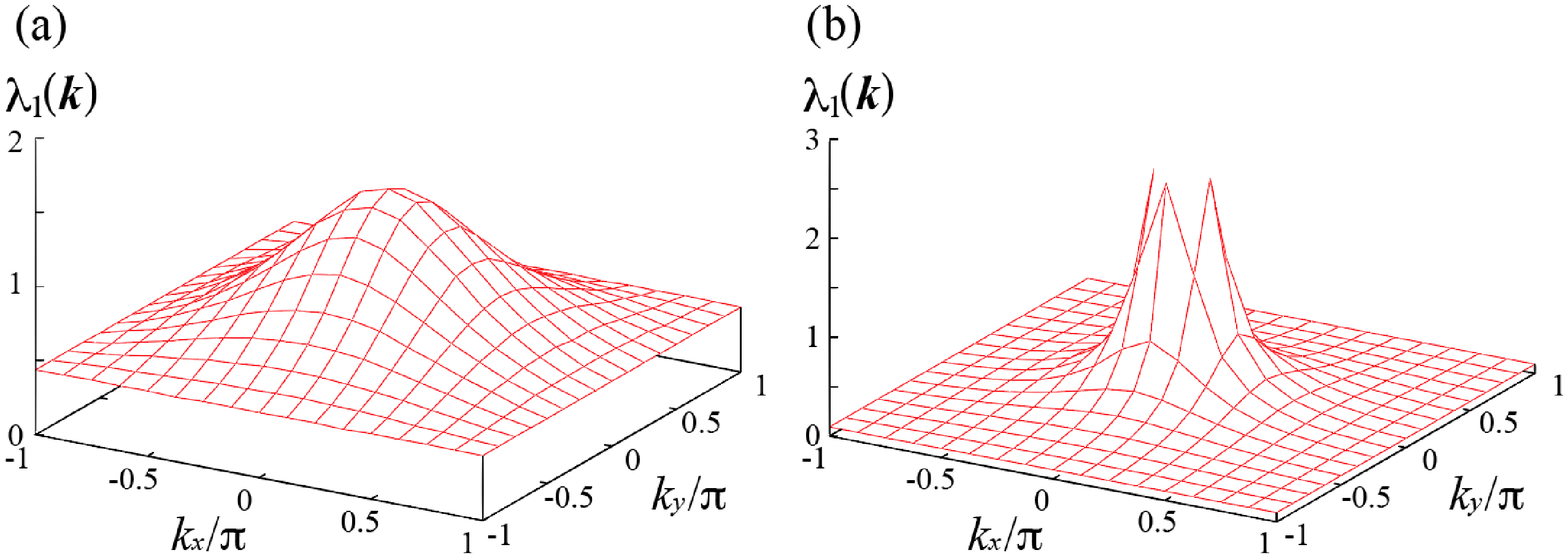}
\end{center}
\caption{(Color online) 
${\bf k} $ dependence of the largest eigenvalue 
$\lambda_1({\bf k})$ in 
$(k_x, k_y, 0)$
plane for $c$=3.0 and $L$=16.
 (a) $T$=2.5 and (b) $T$=0.5. 
The peak value in (b) is $\lambda_1 (\mathbf{0}) \sim 1464 = 0.357 N$.}
\label{fig:Sk}
\end{figure}
%%%%%%%%%%%%%%%%%%%%%%%%%%%%%%%%%%%%%%%%%%%%%%%%%%%%%%%%%

\subsection{Order parameters and critical temperature}

Next, let us investigate spatial correlations.  
The largest eigenvalue $\lambda_1 (\mathbf{k})$ 
of $C_{s \mu , s' \mu' }(\mathbf{k})$ 
is plotted in Fig.~\ref{fig:Sk} 
for the $k_z=0$ plane of the Brillouin zone.  
The peak value grows quickly with lowering $T$, 
and in the ordered phase it is very large 
and proportional to the system size. 
For example, $\lambda_1 (\mathbf{0}) \sim 1464 = 0.357 N$ at $T=0.5$.
This is an evidence that quadrupole moments order 
at this transition.  

At all the temperatures in the calculation, the maximum 
of $\lambda_1 (\mathbf{k})$ in the entire Brillouin zone 
is always located at $\mathbf{k}^*=\mathbf{0}$. 
Of course, this does not mean that the order pattern 
is ferro-type;  
the unit cell of the diamond lattice 
contains $A$- and $B$-sublattices, and the antiferro order does 
not break the translation symmetry.  
To identify the order pattern, 
one needs to analyze the eigenvector of $C_{s \mu , s' \mu'}$. 
Within statistical errors of MC simulation, the 
largest eigenvalue at $\mathbf{k}^*=\mathbf{0}$ 
is doubly degenerate, and its eigenvectors are 
$(x_{Au},x_{Av},x_{Bu},x_{Bv})$$\approx$ 
$(1,0,-1,0)$ and $(0,1,0,-1)$.  
Since the relative sign between the $A$- and 
$B$-sublattice elements is negative, 
the correlation is antiferro 
between the different sublattices.  
The direction of ordered quadrupole needs a more 
careful analysis, and we will study this problem in Sec.~\ref{sec-LT}.  

%%%%%%%%%%%%%%%%%%%%%%%%%%% fig 3 %%%%%%%%%%%%%%%%%%%%%%%
\begin{figure}[t]
\begin{center}
\includegraphics[width=0.45\textwidth]{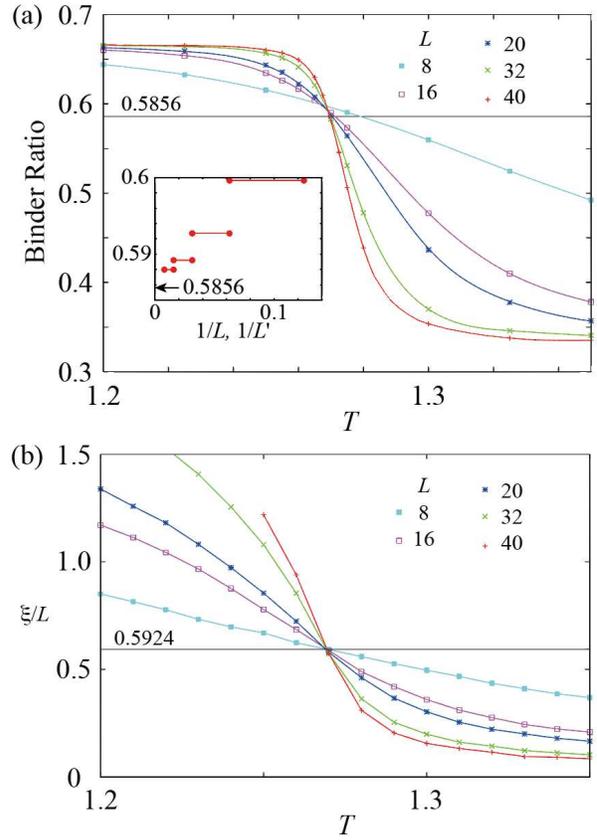}
\end{center}
\caption{ (Color online) 
(a) Binder ratio $\mathcal B$ for $c$=3.0 and $8 \le L \le 40$. 
Lines are a guide for eyes. 
Inset shows the crossing point between the sets of 
two lines for $(L,L')$=(8,16), (16,32), (32,64), and (64,128). 
The two filled circles connected with a line 
indicate $1/L$ and $1/L'$, respectively.
The data for $L\ge 64$ are not shown in the main panel for clarity.
(b) Correlation length $\xi$ for $c$=3.0 and $8 \le L \le 40$.} 
\label{fig:Binder}
\end{figure}
%%%%%%%%%%%%%%%%%%%%%%%%%%%%%%%%%%%%%%%%%%%%%%%%%%%%%%%%%

To determine critical exponents, one has to first locate 
the transition temperature $T_c$ accurately.  
This can be done by calculating the Binder ratio of the primary 
order parameter 
\begin{equation}
{\mathcal B} \equiv 1 - \frac13 
\frac{\bigl\langle ( \mathbf{Q}_{-}^2 )^2\bigr\rangle}
{\bigl\langle \mathbf{Q}_{-}^2 \bigr\rangle^2}.
\end{equation}
Figure \ref{fig:Binder}(a) shows its temperature dependence 
for various system sizes.  
The Binder ratio at $T=T_c$ is a scale invariant quantity, and therefore,  
asymptotically independent of the system size at $T_c$.  
Thus, the crossing point of curves for different $L$'s 
determines the transition temperature $T_c$.  
The value of the Binder ratio at the crossing ${\mathcal B}_c$ is a 
universal quantity, and the known value is ${\mathcal B}_c$=0.5856 
for the XY universality class in the spatial dimension $d=3$.\cite{He4}  
In our results, the crossing points for  large $L$'s approach
the temperature $T_c \approx 1.2695(3)$, and  
extrapolated value of ${\mathcal B}_c$ for $L\to \infty$ is consistent with the known value
as shown in the inset of Fig.~\ref{fig:Binder}(b).  

Another scale invariant quantity is the ratio of two 
length scales, $\xi / L$, and this also determines $T_c$.  
Figure \ref{fig:Binder}(b) shows its temperature 
dependence for the same 
system sizes as for the Binder ratio.  
It is known that $(\xi / L ) \big|_{T_c} = 0.5924$\cite{He4}  
and now this value agrees very well with our data.  
The crossing position leads to $T_c \approx 1.270$, 
which is slightly higher than the estimate based on 
the Binder ratio but agrees with it within statistical 
error in MC simulation.  

%%%%%%%%%%%%%%%%%%%%%%%%%%% fig 4 %%%%%%%%%%%%%%%%%%%%%%%
\begin{figure}[t]
\begin{center}
\includegraphics[width=0.4\textwidth]{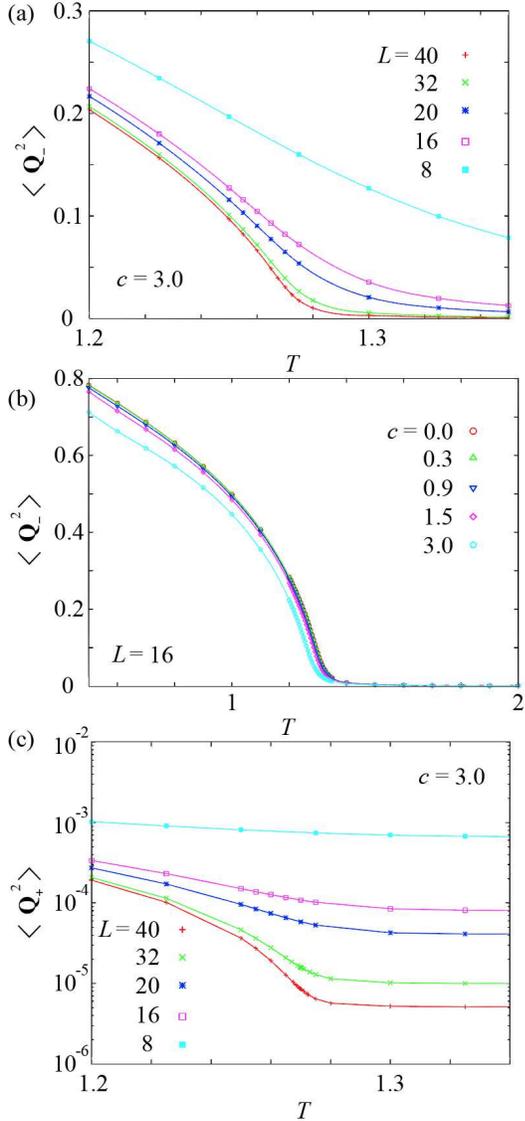}
\end{center}
\caption{ (Color online)
(a) Temperature dependence of 
$\langle \mathbf{Q}_{-}^2\rangle$ 
near $T$=$T_c$ for $c$=3.0 and $8 \le L \le 40$. 
(b) $c$ dependence of 
$\langle \mathbf{Q}_{-}^2\rangle$ for $L$=16. 
(c) Temperature dependence of 
$\langle \mathbf{Q}_{+}^2\rangle$ 
near $T$=$T_c$ for $c$=3.0 and $8 \le L \le 40$. 
}
\label{fig:order}
\end{figure}
%%%%%%%%%%%%%%%%%%%%%%%%%%%%%%%%%%%%%%%%%%%%%%%%%%%%%%%%%

Now, let us see the evolution of order parameters. 
We should note that the ordered phase has two types of 
order parameters and that this is a very characteristic 
point to the antiferro model with the $Z_3$ anisotropy.
The primary one is an antiferro component, and this is 
natural in the antiferro model.  
An interesting one is the secondary order parameter 
and this is a ferro component.  
It arises from the fact that the
order parameter cannot form the complete antiferro. Due to the 
single-ion anisotropy, the antiferro pattern slightly tilts.

Figure \ref{fig:order}(a) shows the temperature dependence 
of the square of the primary order parameter, 
$\langle \mathbf{Q}_-^2 \rangle$.  
With increasing system size, this vanishes above $T_c$, 
but approaches a finite value below $T_c$.  
This also confirms that 
the long range order below $T_c$ is about 
the antiferro alignment of quadrupoles.  
The effects of the $Z_3$ anisotropy are shown in 
Fig.~\ref{fig:order}(b).  
For larger value of the anisotropy parameter $c$, 
the order parameter is slightly reduced and the transition 
temperature also decreases a little, but 
the overall feature does not change.  
The square of the secondary order parameter $\langle \mathbf{Q}_+^2 \rangle$ is shown in Fig.~\ref{fig:order}(c) in 
the log plot. The absolute value is tiny but it clearly develops below $T_c$.

%%%%%%%%%%%%%%%%%%%%%%%%%%% fig 5 %%%%%%%%%%%%%%%%%%%%%%%
\begin{figure}[t]
\begin{center}
\includegraphics[width=0.4\textwidth]{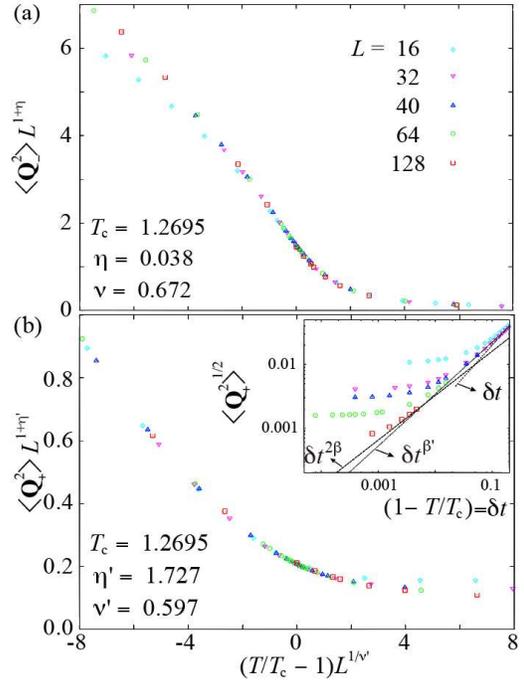}
\end{center}
\caption{ (Color online)
Scaling plot for the two order parameters calculated for 
$16 \le L \le128$ and $c$=3.0. For this analysis, we fix $T_c$=1.2695.
(a) $\langle \mathbf{Q}_-^2 \rangle$ 
and (b) $\langle \mathbf{Q}_+^2 \rangle$. 
Inset: $\delta t$$\equiv$$1-T/T_c$ dependence of uniform moment 
$\langle \mathbf{Q}_+^2\rangle^{1/2}$. 
The three lines $\propto\delta t^{2\beta}, \delta t^{\beta'}$, 
and $\delta t$ with $\beta$ being the order-parameter exponent 
for the $3d$-XY universality class: 
$\beta$=0.3486 and $\beta'$=$\nu'$$(d+\eta'-2)/2$=0.815 
with $d$=3 are also shown for guide to eyes.
}
\label{fig:Qscaling}
\end{figure}
%%%%%%%%%%%%%%%%%%%%%%%%%%%%%%%%%%%%%%%%%%%%%%%%%%%%%%%%%

\subsection{Scaling of primary order parameter}
Now, we examine the critical behavior of primary order parameter in details.  
A conventional analysis is a finite-size scaling, 
and Fig.~\ref{fig:Qscaling} shows the analysis 
for the temperature dependence of order parameter. 
The horizontal axis measures the distance from the critical 
temperature, while the vertical axis is the squared primary 
order parameter, and these two are normalized by the system 
size powered with the constants related 
to the two critical exponents $\nu$ and $\eta$. 
$\nu$ describes divergent behavior of the correlation 
length around the transition temperature, 
$\xi \sim |T - T_c|^{-\nu}$, 
and the anomalous dimension $\eta$ corresponds 
to the power-law exponent of the correlation function 
just at the transition temperature 
$C (\mathbf{k}) \sim 1/|\mathbf{k}|^{2-\eta}$. We have fitted the
numerical data by using different sets of $\nu$ and $\eta$ for the two components. 

For this analysis about the primary order parameter, we have used the values 
known for the three-dimensional ($3d$)-XY universality class, 
$\nu=0.672$ and $\eta=0.038$.  
The data for different system sizes  
nicely fall on a universal curve except for 
the low-temperature region of the smaller 
system sizes as shown in Fig.~\ref{fig:Qscaling}(a).  
Thus, this confirms that the transition belongs 
to the $3d$-XY universality class as long as the primary order
parameter is concerned.

\subsection{Scaling of secondary order parameter}
We have found that the criticality of the ferro secondary order 
parameter is distinctive and very interesting, but previous studies 
by other groups on related models have not addressed this point.  
The presence of the parasitic ferro order was first found in 
our mean-field analysis\cite{Hattori}, and we are going to 
analyze their true criticality based on our 
MC data.  
Let us first explain a natural idea about the expected criticality 
and then show later that our MC results differ from that.  

Naively speaking, the ferro moment is induced by the antiferro 
quadrupole order in the following way.  
In the effective free energy, the secondary order parameter 
couples to the primary order parameter in the lowest order as 
\begin{equation}
\Delta F_{+-} = g_{+-}  [ Q_{u, +}(Q^2_{u,-}-Q^2_{v,-})-2Q_{v,+}Q_{u,-}Q_{v,-}] , 
\end{equation}
with a proper coupling constant $g_{+-}$.  
In the ordered state, the primary order parameter $Q_{v,-}$ 
acquires a finite expectation value
and therefore the coupling term leads to an effective static field 
corresponding to the $Q_{u,+}$ component, 
$h_{u,+} \sim g_{+-} \langle Q_{v,-} \rangle ^2$.  
Thus, it is expected that 
a finite ferro moment is induced in the ordered phase and 
within the linear response region, its size is given by 
$\langle {Q}_{u,+}\rangle \propto \chi_{++}^{uu} h_{u,+} \propto 
\langle {Q}_{v,-} \rangle^2 \propto (1-T/T_c)^{2\beta} $. 
Here, $\chi_{++}^{uu}$ is the quadrupole susceptibility 
of the corresponding ferro component. 
This means that the order parameter critical exponent of the 
ferro component is twice the value of the primary 
order parameter $\beta' = 2 \beta = 0.6972$. 
An important point is that the above scaling presumes 
non-singularity of $\chi_{++}^{uu}$ at the critical point. 
We verified in our previous paper that it is nonsingular within 
the mean-field approximation\cite{Hattori}.

The scaling of the calculated ferro component in our results 
is shown in the inset of Fig.~\ref{fig:Qscaling}(b), 
and one can see that it does not work well with the 
expected exponent $2\beta$. 
We also calculated the ratio 
$[\langle \mathbf{Q}_{+}^2 \rangle ]^{1/2} / \langle \mathbf{Q}_{-}^2 \rangle$, 
and found that it is not independent of $T$ as expected but 
varies strongly with temperature, which also disproves 
the expectation $\beta ' = 2 \beta$.  

In order to determine the value of $\beta'$, we tried a finite 
size scaling for the ferro component and determined the correlation 
length exponent $\nu '$ and also $\eta '$.  
One needs a caution about the meaning of $\eta'$
and it is not clear if $\eta'$ is really the anomalous 
dimension of the ferro component, 
since we do not know if the scaling relation holds 
for $\mathbf{Q}_{+}$.  
The scaled order parameter should be understood as 
$\langle \mathbf{Q}_{+}^2 \rangle L^{2\beta' /\nu' }$, 
and we simply denote the exponent of $L$-dependence $2\beta' /\nu'$ 
by $1+\eta'$. 
We have no prediction on a universality class 
for the ferro secondary order parameter. 
Therefore, no candidate values are available for these exponents, 
and we need an unbiased estimate for their precise values. 
A useful tool for finite size scaling was recently 
developed by Harada\cite{Harada} based on a Bayesian inference analysis. 
We used his method for the ferro components 
and obtained $\nu'=0.597(12)$ and $\eta'=1.727(12)$. 
As shown in Fig.~\ref{fig:Qscaling}(b), the finite size 
scaling works well and the data for different system sizes 
lie on a universal curve.  
The result leads to 
$\beta' = (1 + 1.727 ) \times \nu' /2 \simeq 0.815 > 2 \beta$, 
and 
the inset of Fig.~\ref{fig:Qscaling}(b) shows that this 
value describes nicely the temperature dependence 
in the double-logarithm plot except very close to the critical 
point, where finite size corrections are not negligible.  

It is very interesting that the ferro secondary order parameter 
does not follow the naive scaling with $\beta' =2\beta$ but 
shows another type of criticality with the independent 
exponents $\nu'  < \nu$ and 
$\beta' > 2 \beta$. 
Understanding this criticality 
and the validity of the naive scaling form 
for the secondary order parameter 
is important for the complete 
analysis of the ordered phase, but at the moment we are not 
equipped for reproducing these values by analytical calculations 
like the renormalization group theory.   
From this viewpoint, it is also an important open question 
how to calculate criticality behaviors of non-primary order parameters, 
and we leave this for a future work.

\section{Low-temperature Ordered Phase}
\label{sec-LT}

In this section, we investigate in detail 
the ordered phase in the low-temperature region. 
As explained before, the antiferro 3-state Potts 
model is a much simpler model that has the same symmetry 
with our models; 
each microscopic pseudospin can point to only three directions, 
which corresponds to the limit of 
$c\to \infty$ in the model (\ref{eq-rotor}). 
Although there have been a pile of studies about this problem for 
the Potts case on bipartite lattices \cite{Banavar,Grest,Rosengren,Kolesik,
Pelizzola,Heilmann,Ni,Lapinskas,Rahman}, 
several important points are not settled down 
about this model. 
Most importantly, the nature of the ordered states 
has not been well clarified for the Potts case\cite{Banavar,Grest,Rosengren,Kolesik,Pelizzola,Heilmann,Ni,Lapinskas,Rahman,Okabe}.   
The consensus is that the lowest-temperature phase 
is the broken sublattice-symmetry (BSS) state\cite{Banavar}. 
This is the phase in which the symmetry between the two sublattices is 
broken and this may also be called a ferri state. 
Existence of intermediate-temperature 
phase was also pointed out and an exotic configuration was proposed. 
This is called the permutationally symmetric sublattices (PSS) 
state\cite{Rosengren}. 
There are claims that the high-temperature phase 
undergoes a transition into another intermediate-temperature 
phase named rotationally symmetric (RS) phase. 
Inside this phase the sublattice moments uniformly 
fluctuate\cite{Kolesik,Heilmann}.  

%%%%%%%%%%%%%%%%%%%%%%%%%%% fig 6 %%%%%%%%%%%%%%%%%%%%%%%
\begin{figure}[t]
\begin{center}
\includegraphics[width=0.4\textwidth]{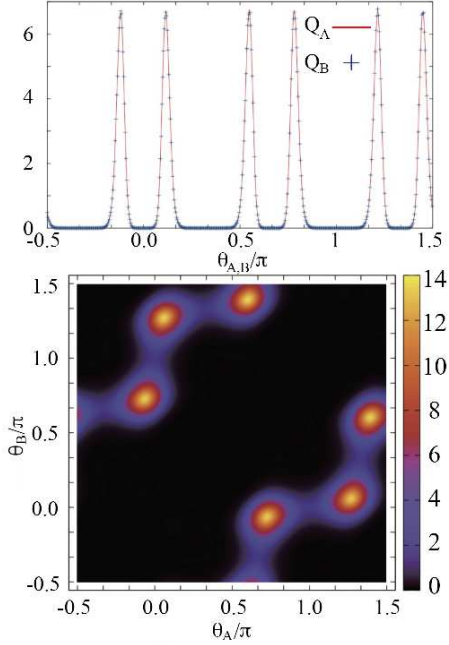}
\end{center}
\caption{ (Color online)
(a) Distribution of $\mathbf{Q}_s (s=A,B)$ as a function of 
$\theta$ for $L$=16.  
(b) The nearest-neighbor correlations of $\mathbf{Q}_A$ and 
$\mathbf{Q}_B$ for $L$=16.  
Temperature is $T$=0.5 
and $c=0.5$ in both data.}
\label{fig:Qdist}
\end{figure}
%%%%%%%%%%%%%%%%%%%%%%%%%%%%%%%%%%%%%%%%%%%%%%%%%%%%%%%%%

We will examine the following 
two points in 
our rotor model (\ref{eq-rotor}).  
The first one is about the nature of the low-temperature 
ordered phase. 
The second one is whether an intermediate temperature phase exists.  

To investigate the pseudospin configuration in the 
ordered phase, we first calculate the distribution of macroscopic 
sublattice moments, $\mathbf{Q}_{A}$ and $\mathbf{Q}_{B}$ 
[see eq.~(\ref{eq3:Q})], 
in our MC simulations.  
Figure \ref{fig:Qdist}(a) shows the distribution 
below $T_c$ of their directions, 
$\mathbf{Q}_s \propto (\cos \theta_s , \sin \theta_s )$. 
While the distribution is almost independent of direction $\theta_s$ 
above $T_c$, 
each sublattice moment in the ordered phase mainly 
points around 6 directions as shown in Fig. \ref{fig:Qdist}(a). 
The distribution is identical for the two sublattices.  
These 6 directions are not equally separated on 
the circle in the $\mathbf{Q}$-space, but located on both sides of the 
three favored directions of the $Z_3$ anisotropy.  
This is consistent with our previous 
conclusion based on the mean-field analysis \cite{Hattori} that 
the ordered phase has a 6-fold degeneracy.  
Due to global updates in our simulation, 
the order parameter migrates from one of the 6 stable 
points to another, and the 6 points become equally 
populated after long Monte-Carlo runs in our simulation.  

Figure \ref{fig:Qdist}(a) shows that 
$\mathbf{Q}_A$ and $\mathbf{Q}_B$ have the same distribution, 
but this does not mean that they point to the same direction.   
We have examined the correlation between $\mathbf{Q}_A$ and $\mathbf{Q}_B$ 
by calculating the two-body distribution of the directions of 
microscopic pseudospins on nearest-neighbor site pairs, 
$(\theta_A (\mathbf{r}), \theta_B (\mathbf{r}'))$.  
The distribution is calculated by evaluating nearest-neighbor 
configuration for all the sites during MC  
steps at $T=0.5$, and its value is shown with 
color plot in Fig.~\ref{fig:Qdist}(b).  
This plot provides valuable information on the moment 
configuration in the ordered phase.  
There are 6 peaks, and the most important point is that 
the 6 values of $\theta_s$ are different to each other 
for either of $s=A$ and $B$. 
This manifests that for each direction of ordered quadrupole 
in the $A$-sublattice, the favorite direction on the $B$-sublattice 
neighbor sites is \textit{uniquely} determined, and \textit{vice versa}.  
Therefore, the spatial quadrupole configuration is 
uniquely determined in the ordered state except for a domain structure 
due to the 6-fold degeneracy related to the global $Z_3$ symmetry.  

This conclusion differs from the case of the 3-state Potts 
model.  
It is believed that  its lowest-temperature 
region is the phase of the broken sublattice-symmetry (BSS) 
state.\cite{Banavar}  
In this state, Potts spins on the $A$-sublattice point 
to one of the three directions, while $B$-sublattice 
Potts spins point to the two other directions 
randomly from site to site and this yields 
$\frac12 \log 2$ residual entropy at $T=0$. 
Alternatively, the $A$- and $B$-sublattices switch their 
roles. 
Therefore, the symmetry between the two sublattices is 
broken. 
This may also be called a ferri state, since 
$\langle \mathbf{Q}_{s} \rangle$=
$-2 \langle \mathbf{Q}_{s'} \rangle \ne \mathbf{0}$.  

It has been also proposed that an intermediate 
phase exists between the disordered phase and the low-temperature 
ordered phase, and that it is an exotic one named 
the permutationally symmetric sublattices (PSS) state.\cite{Rosengren} 
In this phase, the most-favorite quadrupole (Potts spin) direction in the $A$-sublattice 
is the least-favorite direction in the $B$-sublattice, 
and \textit{vice versa}.  
We find that the low-temperature phase in our model is 
equivalent to the PSS state, 
and show this below by calculating sublattice moments. 

Let $p_n (s)$ be the probability that $s$-sublattice 
Potts ``spins'' point to the direction $\theta_n  \equiv 2n\pi/3$ 
($n \in \{ 0,1,2 \}$).  In one of the PSS states,
$ p_2 (A)=p_1 (B) = \frac13 ( 1 - w - w')$, 
$ p_0 (A)=p_0 (B) = \frac13 (1 - w)$, 
and $ p_1 (A) = p_2 (B) = \frac13 (1 + 2 w + w' )$, 
with $ w, w' > 0$.  
For this case, the sublattice moment averages are 
\begin{eqnarray}
&&\langle \mathbf{Q}_A \rangle 
= w ( \cos \theta_1 , \sin \theta_1 ) +  \frac{1}{\sqrt{3}} w' (0, 1) , 
\nonumber\\
&&\langle \mathbf{Q}_B \rangle 
= w ( \cos \theta_2 , \sin \theta_2 ) -  \frac{1}{\sqrt{3}} w' (0, 1).
\label{eq:SPP}
\end{eqnarray}
Their directions are slightly shifted from the directions 
of Potts spins, 
$\theta_A = \theta_1 - \delta \theta$ and 
$\theta_B = \theta_2 + \delta \theta$, 
where 
$\tan \delta \theta =  w' / [{\sqrt{3}}
( 2w + w' )]$.
This is exactly what is realized in our simulations.

It has been also claimed that the high-temperature phase 
undergoes a transition into another intermediate-temperature 
phase when the model is on a simple cubic lattice.\cite{Kolesik,Heilmann}   
This is called the rotationally symmetric (RS) phase, 
and within this phase the moments in each sublattice fluctuate 
their directions uniformly in the entire range of angle  
$ 0 \le \theta_s < 2\pi$.\cite{Kolesik,Heilmann}  
If this phase is realized, the two-angle distribution
in Fig.~\ref{fig:Qdist}(b) 
should have shown two straight bright bands 
that are parallel to each other and separated 
by the relative angle difference $\pi$.  
We note that this is the behavior expected for the completely isotropic 
XY antiferromagnet, and also in the low-temperature part 
of the disordered phase in our models.  
Almost the same behavior is observed above $T_c$ in our simulation, 
although the bands are not completely straight but 
slightly wind due to the $Z_3$ anisotropy.  

Thus, the two-angle distribution in Fig.~\ref{fig:Qdist}(b) 
in our calculation negates the BSS state and also the RS state. 
The results agrees only with the distribution in the PSS state.  
Rahman, {\it et al}.,\cite{Rahman} demonstrated that the PSS state claimed in Ref. \citenum{Rosengren} 
for the AF 3-state Potts model on a simple cubic lattice is an artifact of their incorrect way when taking the sublattice
degeneracy into account. For a diamond lattice model,
Ref. \citenum{Lapinskas} 
suggests the PSS state occurs, which is 
consistent with our results in this paper. We note that in our simulations there is only one transition, and thus, there is no intermediate phase in our model.

\section{Temperature-Magnetic Field Phase Diagram} 
\label{sec-THphase}

So far, we have discussed the critical properties of the model 
(\ref{eq-rotor}) with no external field. 
Since quadrupole moments are equivalent to a second-order product of 
magnetic dipoles, 
a magnetic field ${\bf H}$ leads to a quadratic Zeeman coupling of 
$\mathbf{Q}_i$.  
Various quadrupolar systems such as Pr-based 1-2-20 \cite{Onimaru0,Onimaru,Onimaru2,Ishii1,Ishii2}
and PrPb$_3$ \cite{PrPb3}, have several ordered phases in magnetic fields, 
and this depends on the field direction.  
Thus, it is worth examining the effects of magnetic fields when 
 the quadratic Zeeman coupling is taken into account.\cite{Hattori}  
We denote this coupling as 
\begin{equation}
   H_{\rm mag}=-h \sum_i Q_i\cos\theta_i, \label{Hmag}
\end{equation} 
where 
$h\propto 2H_z^2 - H_x^2 - H_y^2$ is the conjugate field of $Q_u$ component, 
and the sign of $h$ depends on the direction 
of ${\bf H}$ and the microscopic parameters in the system such as the
 CEF level structures\cite{Hattori}. 
Magnetic field influences not only the direction but also the amplitude 
of quadrupole moments\cite{Hattori}.  
Their amplitude may vary from site to site, as was shown
in our previous study\cite{Hattori}, and an experiment also found this
in the related material PrPb$_3$\cite{PrPb3}.  In order to take account of
this effect, we will use  throughout this section the model
(\ref{Model})-(\ref{Hint}), in which $|{\bf Q}_i|$'s fluctuate.

As for the relation between the conjugate field $h$ and the real magnetic field ${\bf H}$, for example, 
$h>0$ corresponds to ${\bf H}\parallel$ [001] and $h<0$ to ${\bf H}\parallel$ [110].
For other magnetic-field directions such as [010] or [100], 
$H_{\rm mag}$ contains an additional term 
$-h' \sum Q_i \sin \theta_i$, where $h' \propto \sqrt{3} (H_x^2 - H_y^2)$.  
For ${\bf H}\parallel [111]$, $h$ and $h'$ are both zero and the coupling 
to quadrupole moments vanishes.  
In this case, the leading-order effect of magnetic field is 
a coupling to octupole moments, but this is beyond the scope of the present study 
and we will not discuss it in this paper.

\subsection{Symmetry of the quadrupole model in magnetic field} 

Before showing the results of phase diagram, let us 
briefly explain the symmetries of the Hamiltonian 
including the quadratic Zeeman coupling $H_{\mathrm{mag}}$.  
This will be important for discussing phase transitions 
in magnetic fields.  

In the case of no magnetic field, $h=h'=0$, the Hamiltonian 
has three types of symmetries.  
The first is the time-reversal symmetry, and the second 
is the $Z_2$ symmetry of the two-sublattice exchange.
The last one is related to the $Z_3$ anisotropy, and 
its precise symmetry is the symmetric group $S_3$ related to 
permutations of the three special $\mathbf{Q}$ directions. 
Recall that these directions correspond to the $x$, $y$, and 
$z$-axes of the cubic lattice structure in the real space.  
Since our models are about continuous ``$\phi^4$ spins'', 
more precisely speaking, the symmetry operations 
in the $\mathbf{Q}$ space are the three rotations 
with angle $\theta=0,\pm 2\pi /3$ and the three mirrors 
like $R_v : Q_v \rightarrow - Q_{v}$.  
By using the point group nomenclature, 
this symmetry group is $D_3$ or $C_{3v}$, which are 
isomorphic to $S_3$.  

When a finite uniform magnetic field $\mathbf{H}$ 
is applied, the $Z_2$ sublattice symmetry persists. 
The time reversal symmetry also persists in our 
models of quadrupole moments, since 
the quadrupole operators and their conjugate field are 
both invariant.  
The internal $D_3$ symmetry is generally broken, 
because the three principle axes in the lattice 
structure are no longer equivalent due to the applied field.  
However, the exception is the case of $\mathbf{H} \parallel$ 
[001], [110], or their equivalent directions.  
In this case, two of the three axes in the lattice 
remain equivalent, and correspondingly 
the $D_3$ symmetry is not completely 
broken but reduced to the dihedral symmetry.  

The internal symmetry is directly related to the 
point group symmetry of the lattice structure.  
The above symmetry operations in the internal $\mathbf{Q}$ 
space are equivalent to some lattice rotations.  
The $\pm 2\pi /3$ rotation in the $\mathbf{Q}$ space 
is also a three-fold rotation about one of the trigonal 
axes of the lattice, {\it e.g.}, [111] direction.  
The mirrors in the $\mathbf{Q}$ space are 
$\pm \pi /2$ rotation about one of the $x$, $y$, and $z$ axes 
of the lattice.  
For example, $R_v$ is equivalent to $\pm \pi /2$ rotation 
about the $z$ axis.  

Thus, the symmetries of the Hamiltonian under the field $h$ 
are described by the group of symmetry operations, 
$G = \{ 1, P_{AB}, R_v , P_{AB} R_v  \}$, where $P_{AB}$ 
denotes the exchange of the two sublattices, and 
note that $P_{AB}^2 = R_v^2 =1$ and 
$P_{AB}  R_v = R_{v} P_{AB}$.  
Since this is an Abelian group, all of the four irreducible 
representations are one-dimensional, and this means 
that the order parameters of spontaneous symmetry breaking 
are scalars and the universality class of the corresponding 
phase transitions is the $3d$-Ising universality class, except at multi-critical 
points.

%%%%%%%%%%%%%%%%%%%%%%%%%%% fig 7 %%%%%%%%%%%%%%%%%%%%%%%
\begin{figure}[t!]
\begin{center}
\includegraphics[width=0.5\textwidth]{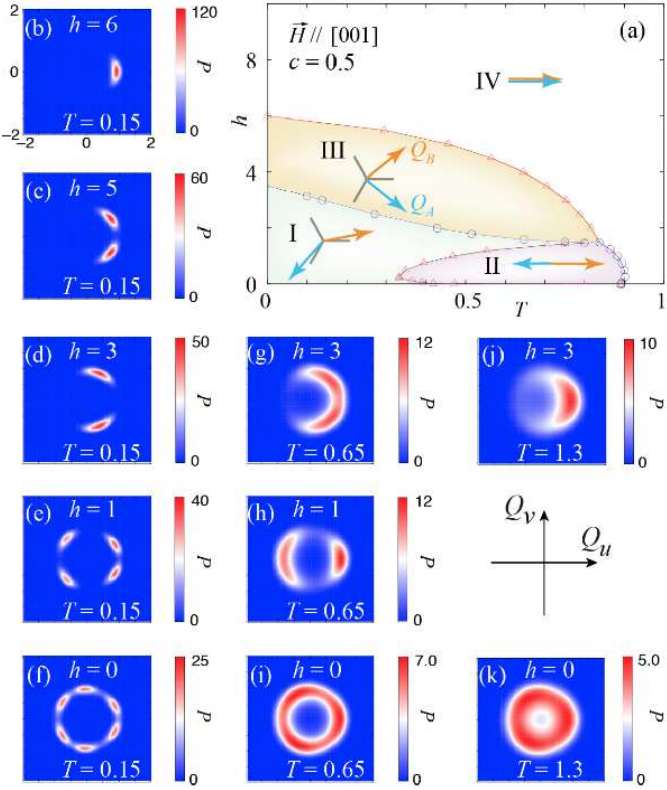}
\end{center}

\caption{
(Color online) 
(a) $T$-$h$ phase diagram for the anisotropy $c=0.5$. 
The conjugate field is proportional to the square of magnetic field, 
$h \propto H_z^2$.  
Quadrupole configuration in each phase is schematically depicted with arrows 
with the 3-fold axes for the phases I and III. 
Phase I has four stable domains; the one with the depicted configuration 
and the other domains obtained with transformation $R_v$ and/or $P_{AB}$. 
[See (e)].   
Just on the $h=0$ line only one ordered phase I exists, and 
this has two additional stable domains that are transformed from 
the depicted one by $\pm 2\pi /3$ rotation in the $\mathbf{Q}$ space.  
[See (f) and (i)].  
(b)-(k) Distribution of local quadrupole moments $P(Q_u,Q_v)$ 
for typical values of $T$ and $h$ calculated in the system of $L=16$.  
The part of $-2 \le Q_u, Q_v \le 2$ is shown in all the panels.  
In the ordered phases I-III, different peaks correspond to different domains.  
The higher temperature result (i) in the phase I has broadened peaks that 
nearly merge pairwise due to large thermal fluctuations.} 
\label{fig-HT}
\end{figure}
%%%%%%%%%%%%%%%%%%%%%%%%%%%%%%%%%%%%%%%%%%%%%%%%%%%%%%%%%%%%%%%%%%%%%%%%%%%%%%%

\subsection{$T$-$h$ phase diagram for $h>0$}
\label{Thphase-positive}

Let us first discuss the $h>0$ part of the $T$-$h$ phase diagram. 
This part corresponds the case of magnetic field $\mathbf{H} \parallel$ [001], 
and $h \propto H_z^2$.  
The Hamiltonian has two types of symmetries, 
and ordered phases are related to their breaking.  
One symmetry operation is the exchange of the two sublattices $P_{AB}$, 
and the other is the mirror that reverses the $v$-component
$R_v : Q_v \rightarrow -Q_v$. 
In terms of the point group, the latter is  
the diagonal mirror operation, $(x,y,z) \rightarrow (y,x,z)$.  
Since both of $P_{AB}^2$ and $R_v^2$ are the identity operator, 
the Hamiltonian has corresponding parity symmetries, namely 
$Z_2 \otimes Z_2$ symmetry, when the quadratic Zeeman coupling (\ref{Hmag}) 
is present.  
As will be shown later, it is more convenient to consider 
instead of $P_{AB}$ its product with the mirror operation, 
$R_{AB} \equiv P_{AB} R_v$, and this also has a parity character, 
$R_{AB}^2 =1$.  
Ordered phases in the magnetic field along the [001] direction 
are related to how this $Z_2 \otimes Z_2 $ symmetry is broken.  

\begin{figure}[t]
\begin{center}
\includegraphics[width=0.48\textwidth]{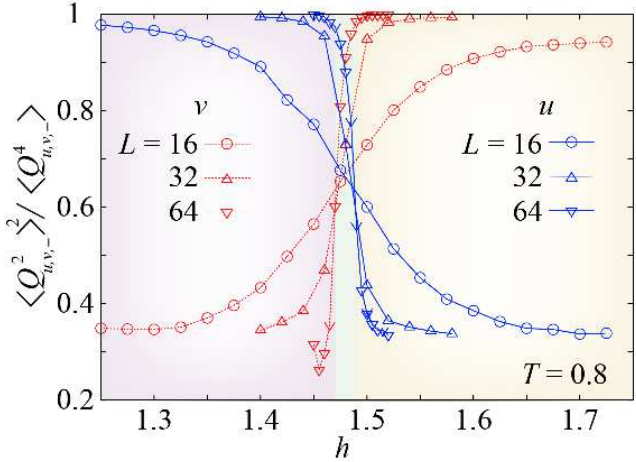}
\end{center}
\caption{
(Color online) 
Binder Ratio 
${\mathcal B}_u$=%
$\langle Q_{u,-}^2\rangle^2/\langle Q_{u,-}^4\rangle$ 
(dotted lines) and 
${\mathcal B}_v$=%
$\langle Q_{v,-}^2\rangle^2/\langle Q_{v,-}^4\rangle$ 
(full lines) 
for $T$=0.8 and $L$=16 ($\bigcirc$), $L$=32 ($\triangle$), and 
$L$=64 ($\bigtriangledown$).}
\label{fig-boundary}
\end{figure}

We determined the phase diagram by the MC calculations 
performed with global updates modified for the $R_v$ symmetry,\cite{Wolff}  
and the result is shown in Fig.~\ref{fig-HT}(a) for the anisotropy $c=0.5$.  
The phase boundaries are determined by calculating Binder ratio for each 
order parameter with typically $L \le 64$ (see also Fig.~\ref{fig-boundary} and discussions there). 
Three ordered phases appear and 
their symmetry is summarized in Table \ref{tbl_symmetry}.  
All the three ordered phases were found 
in our previous mean-field study \cite{Hattori}, but 
the phase boundaries have a different geometry and different shapes.  

The part at low-$T$ and low-$h$ is the phase I 
and this is 
essentially the same as 
the ordered phase at $h=0$.  
The difference from the $h=0$ case is about the degeneracy of stable domains.  
While the $h=0$ ordered phase has six types of domains, two of the six 
become unstable for $h>0$, and the stable four are related 
with each other under $R_v$, $R_{AB}$, and $R_{AB}R_v$ operations.  

In a higher field part, there appears the phase III and the quadrupole moments 
exhibit a canted configuration; 
$Q_{v,-}\ne 0$ and $Q_{u,+}>0$. 
The high-$T$ side of the phase I touches another ordered phase, 
the phase II, and the quadrupole moments show a collinear (ferri) configuration 
there; $Q_{u,\pm}\ne 0$ and $Q_{v,\pm}=0$.
The polarized phase IV is the part at very large $h$, 
and the pseudospins are all aligned in the direction of the conjugate field, 
$Q_{u,+}>0$.  
This phase smoothly continues to the disordered phase at $h=0$ 
for high temperatures $T>T_c=0.8914(1)$. 
All the transitions are the second order for $c=0.5$.  

The phase I is stabilized by the inter-site interaction $J$, and this 
explains why it appears at low fields. 
The phase III has a configuration of symmetrically canted moments, 
and this gains an energy from both of the $J$ and the $h$ terms.  
The phase II has a collinear configuration of quadrupoles.  
It is not straightforward to understand its stability, 
but it is likely that it is stabilized by thermal fluctuations.  
The collinear configuration has a larger number of low-energy excited 
states compared with other orderings such as non-collinear configurations, 
and the corresponding large entropy lowers the free energy when the temperature 
is not so low. 
It should be noted that the phase II is very sensitive to the third-order anisotropy term $c$.  
At $c$=0, this phase vanishes, and 
its region grows with increasing $|c|$. Its detailed $c$ dependence 
will be discussed at the end of this subsection.

%%%%%%%%%%%%%%%%%%%%%%%%%%% table 1 %%%%%%%%%%%%%%%%%%%%%%%
\begin{table*}[!t]
\begin{center}
\caption{
Symmetry and order parameters in the phases I-IV and I$'$.  
Each symmetry operation is marked ``\textit{inv.}'' 
if the phase is invariant with its operation, 
or ``$\times$'' otherwise.  
Note that $R_{AB} R_v $ equals the simple sublattice exchange $P_{AB}$.  
I-III are the ordered phases in the magnetic field 
$\mathbf{H} \parallel$ [001], 
while I$'$ is the ordered phase in  $\mathbf{H} \parallel$ [110].}
\vspace{3mm} 
\begin{tabular}{lccccccc}
\hline\hline
 &  $R_{v}$ & $R_{AB} $ & $R_{AB} R_v $ & 
$ \langle Q_{u,+} \rangle $ & 
$ \langle Q_{u,-} \rangle $ & 
$ \langle Q_{v,+} \rangle $ & 
$ \langle Q_{v,-} \rangle $ \\
\hline
\ Phase I & $\times$ & $\times$ & $\times$ & 
 $\ge 0$ & $ \ne 0$ & $\ne 0$ & $\ne 0$ 
\\
\ Phase II & \textit{inv.} & $\times$ & $\times$ & 
 $>0$ & $ \ne 0$ & $0$ & $0$ 
 \\
\ Phase III & $\times$ & \textit{inv.} & $\times$ & 
 $>0$ & $0$ & $0$ & $\ne 0$ 
\\
\ Phase IV & \textit{inv.} & \textit{inv.} & \textit{inv.} & 
 $\ne 0$ & $0$ & $0$ & $0$ 
 \\
         \hline
\ Phase I$'$ & $\times$ & \textit{inv.} &  $\times$ &
 $<0$ & $0$ & $0$ & $\ne 0$ 
\\ 
         \hline\hline
\end{tabular}
\label{tbl_symmetry}
\end{center}
\end{table*}
%%%%%%%%%%%%%%%%%%%%%%%%%%%%%%%%%%%%%%%%%%%%%%%%%%%%%%%%%%%

{\it Quadrupole order---}
Now, we investigate a microscopic structure of the order parameter 
in these phases 
by analyzing the distribution of local quadrupole moments.  
Figures \ref{fig-HT}(b)-(k) show the distribution 
of the $A$-sublattice quadrupole moments 
\begin{eqnarray} 
  P(\mathbf{Q})=
  \frac{2}{N}\sum_{i\in A} 
  \left\langle \delta(\mathbf{Q}-\mathbf{Q}_i)\right\rangle , 
\end{eqnarray} 
where $\langle \cdots \rangle$ denotes the MC average, and 
this is normalized such as $\int d\mathbf{Q} P(\mathbf{Q})=1$. 
In our simulations, the configuration migrates all the equivalent 
domains (if present) with the help of implemented global updates\cite{Wolff}.  
This means that the distribution is invariant for all the symmetry 
operations of the model, and one symmetry is about sublattice exchange.  
Therefore, the distribution in the $B$-sublattice 
is identical to the above one within statistical errors.  

The distribution $P(\mathbf{Q})$ is plotted in Figs.~\ref{fig-HT}(b)-(k) 
for several typical points in the $T$-$h$ parameter space, and 
it clearly exhibits the characteristics of each phase.  
The point (e) is in the phase I, while (c), (d) and (g) in the phase III.  
(h) is in the phase II, and the ferri nature is visible there.  
(b) and (j) are in the polarized phase IV, and 
the distribution has only one peak there.  
Note that on the $h=0$ line, the $Z_3$ symmetry in the $\mathbf{Q}$-space 
is recovered as it should be.   
At low temperatures, the distribution shows well-separated peaks and 
they correspond to different stable domains in the ordered phases.  
The sequence of (b)$\to$(e) clearly demonstrates 
how the canted antiferro quadrupole order 
develops with decreasing $h$ from the polarized phase IV.
With the information of $P(\mathbf{Q})$ alone it is not conclusive that the
order parameters have those configurations depicted in Fig.~\ref{fig-HT}(a). 
Thus, we have additionally calculated the nearest-neighbor correlation, as was done
in Sec. \ref{sec-LT} and 
confirmed these configurations. 
 The point of $h=0$ and $T=0.65$ is inside the phase I 
and this phase has six domains.  However, the distribution 
at this point is very broadened as shown in Fig.~{\ref{fig-HT}}(i), 
and each pair of nearest peaks is indistinguishable due to
large thermal fluctuations.  Lowering temperature suppresses these thermal
fluctuations and each pair evolves into well-separated spots 
as shown in (f). 
The origin of large thermal fluctuations is related to the 
peculiar shape of the boundary between the phases I and II,  
and we will briefly discuss this below.

{\it Phase boundaries---} 
Now, we discuss the phase boundaries. 
The present MC results differ from our previous mean-field analysis\cite{Hattori} in several important points.  
In the mean-field analysis,  the phase II lays between the phases I 
and III and the phase I does not 
touch the phase III.   
This is one important difference from the result in Fig.~\ref{fig-HT}(a).  
Another but related difference is the boundary between the phases I and II.  
It did not show a reentrant behavior, and 
the phase II extended down to the $T=0$ limit in Ref. \citenum{Hattori}.  

Before discussing the difference from the mean-field result, 
let us examine the order of transitions based on symmetry arguments.  
In the mean-field phase diagram, 
the transition from the phase II to the lower-field phase I is  
second order, while the transition to the higher-field phase III is 
first order.
The symmetry of the ordered phases can explain 
the different orders of the transitions in our previous 
phase diagram and also the newly found I-III phase boundary.   
First of all, the phase I has no symmetry and 
both of $R_{v}$ and $R_{AB}$ symmetries are broken.  
In the phase III the $R_{AB}$ symmetry remains unbroken.  
Therefore, the transition to the phase I is related to breaking 
the $R_{AB}$ symmetry, and this is expected second order.  
The same is true for the I-II transition. 
It is related to breaking the remaining $R_v$ symmetry,  
and thus, expected second order.  
The phases II and III have different types of order parameters, 
and therefore, their phase transition should be first order.  
Orders of the transitions have been successfully explained 
for the mean-field and MC phase diagrams.  

\begin{figure}[t]
\begin{center}
\includegraphics[width=0.4\textwidth]{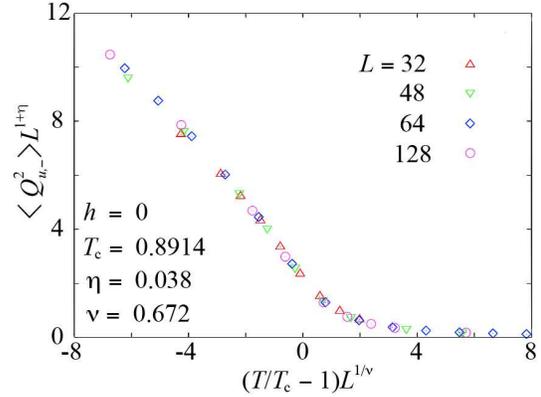}
\end{center}
\caption{
(Color online) 
Scaling plot of $\langle Q_{u,-}^2\rangle$ for $h=0$, $c$=0.5 
and $32 \le L \le 128$.} %and (b) $c=3$.}
\label{fig-scalingH0soft}
\end{figure}

Let us now analyze the difference in the topology of 
the ordered phases between the mean-field and MC studies.  
First, in the MC phase diagram, 
it should be noted that the lower-field side of the I-II phase boundary 
continues to the zero field critical point $(T,h)=(T_c (h=0),0)$ 
without crossing or touching the $h=0$ line before that, 
but the part of $0.4 \simle T < T_c (h=0)$ is so close to the $h=0$ line.  
Calculation to show this needs extremely high precision, and 
it is not easy to directly check this point.  
Therefore, we try an alternative proof.  
Using the MC data obtained in Sec.~\ref{sec-zeromag},  
we can show that the transition at $T_c (h=0)$ upon varying 
temperature is surely of the $3d$-XY type. 
This ensures that both $Q_u$ and $Q_v$ fluctuations diverge 
at this critical point, which in turn ensures the phase boundary 
between I and II continues up to $T_c (h=0)$.  
To confirm this universality class, we carried out the finite-size 
scaling analysis of the fluctuation of the primary order parameter 
$\langle Q_{u-}^2 \rangle$,
and show the result in Fig.~\ref{fig-scalingH0soft} 
analyzed with the known exponents of the $3d$-XY universality class.  
The data for different system sizes nicely collapse onto a universal 
curve and this confirms the transition is of the $3d$-XY type as expected.

\begin{figure}[t]
\begin{center}
\includegraphics[width=0.48\textwidth]{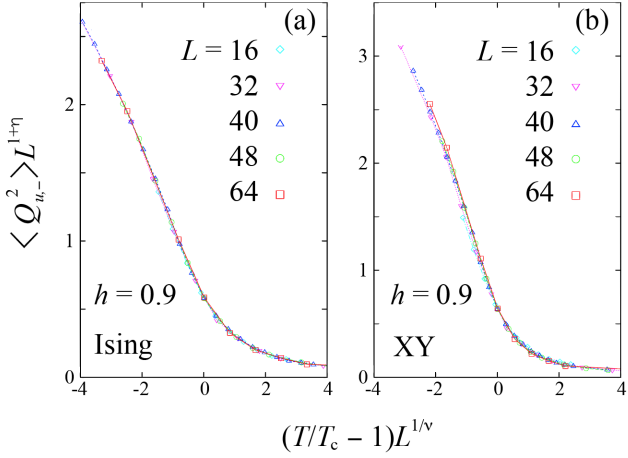}
\end{center}
\caption{
(Color online) 
Scaling plot of $\langle Q_{u,-}^2\rangle$ for $h$=0.9, $c$=0.5, 
$T_c$=0.88598(5), and $16 \le L \le 64$. 
Comparison between 
(a) $3d$-Ising ($\nu$=0.62999 and $\eta$=0.03631)\cite{CFBootstrap} and 
(b) $3d$-XY ($\nu$=0.672 and $\eta$=0.038). } %and (b) $c=3$.}
\label{fig-scalingH0.9soft}
\end{figure}

Applying the magnetic field reduces the symmetry of the system and 
the $Z_3$ symmetry (more precisely, $D_3$ symmetry) is lost for $h\ne 0$.
As explained before, 
the expected universality class of the transition between the para and 
ordered phases (II and III) 
is the $3d$-Ising class. 
We have checked this point by analyzing the transition at fixed $h=0.9$, 
where the transition temperature $T_c (h=0.9)=0.88598(5)$. 
Figure \ref{fig-scalingH0.9soft} shows the scaling plot of 
the primary order parameter $Q_{u,-}$. 
To identify the transition, two universality classes are examined.  
One is the $3d$-Ising class shown in the panel (a) and the other 
is the $3d$-XY class in (b). 
Since the two sets of the exponents $\nu$ and $\eta$ are very similar, 
the difference is small but one can see that the scaling with 
the $3d$-Ising exponents\cite{CFBootstrap} works better for a wider range of the reduced 
temperature.  

Let us also examine other parts in the phase diagram; 
the I-III phase boundary and the high-field side of the I-II boundary.   
An important difference from our previous study\cite{Hattori} 
is that the bicritical point of the phases II, III, and IV 
in the mean-field phase diagram is now replaced 
by a tetracritical point where all the phases I-IV meet. 
Our MC calculations indicate that the phases II and III touch 
only at one point and the phase I intervenes between them. 
To confirm this, we have demonstrated the presence of 
two different transitions when $h$ varies at a fixed temperature 
near the tetracritical point.  
This is done by calculating $h$-dependence of characteristic 
quantities of the two transitions at $T=0.8$ fixed.  
Figure \ref{fig-boundary} shows the two Binder ratios 
${\mathcal B}_{\mu}$=$\langle Q_{\mu,-}^2\rangle^2/\langle Q_{\mu,-}^4 \rangle$ 
$(\mu=u,v)$ for $L$=16, 32, and $64$ as a function of $h$.  
${\mathcal B}_u$ becomes $L$ independent at the I-III phase boundary, 
while ${\mathcal B}_v$ does at the I-II phase boundary.  
Although the convergence with $L$ is not sufficient, 
the tendency clearly shows that the two transitions occur 
at different $h$ values.

\begin{figure}[t]
%\vspace{-0.5cm}
\begin{center}
\includegraphics[width=0.5\textwidth]{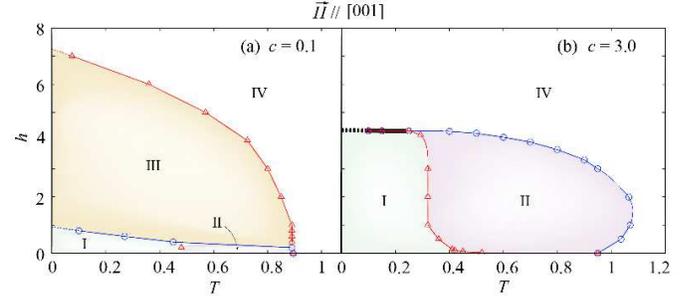}
\end{center}
\caption{
(Color online) 
$T$-$h$ phase diagram for 
(a) $c$=0.1 and (b) $c$=3.0 under magnetic field ${\bf H} \parallel [001]$. 
In (b), the part of thick line shows the first-order 
transition between the phases I and IV. 
The dotted line is the extrapolation to $T$=0.} 
\label{fig-c-dep}
\end{figure}

Finally, let us discuss the dependence of the phase diagrams 
on anisotropy strength $c$.
Figure \ref{fig-c-dep} shows the $T$-$h$ phase diagrams for 
$c$=0.1 and 3.0.
Compared to the case of $c$=0.5 in Fig.~\ref{fig-HT}(a), 
the phase II significantly shrinks at the smaller $c$ 
[Fig.~\ref{fig-c-dep} (a)],  
while grows at the larger $c$ [Fig.~\ref{fig-c-dep} (b)].
Thus, the anisotropy $c$ stabilizes the phase II.  
We should also mention that the phase III disappears 
for larger $|c|$ ($c=3.0$) and the phase I now directly 
touches the phase IV at high field.
Within our MC simulations, the IV-I phase boundary is a line of 
first-order transition. 
We can explain this result based on a symmetry argument. 
The transition from the disordered phase IV to the ordered phase I 
in the conjugate field $h$ requires that the two symmetries 
$R_v$ and $R_{AB}$ need to be broken at the same time.  
The system has no further symmetry to enforce this, and 
therefore if these two phases share a boundary of a finite length, 
the transition should be first order.
At the end point of the first-order transition line, 
the three phases II-IV meet and therefore this is a bicritical point. 
Needless to say, it does not refute the presence of 
the tetracritical point discussed before for smaller $c$, since 
the phases IV and I share only one point there.  
Other boundaries are lines of second order 
transitions as in the case $c=0.5$.  
We also note that the reentrant behavior in the low-field part 
of the phase diagram is greatly enhanced at the larger anisotropy $c$=3.0.  
The reentrant behavior will be discussed in 
details in Sec.~\ref{sec-Discussion}

%%%%%%%%%%%%%%%%%%%%%%%%%%% fig 8 %%%%%%%%%%%%%%%%%%%%%%%
%%%%%%%%%%%%%%%%%%%%%%%%%%%%%%%%%%%%%%%%%%%%%%%%%%%%%%%%%%

%%%%%%%%%%%%%%%%%%%%%%%%%%% fig 9 %%%%%%%%%%%%%%%%%%%%%%%
%%%%%%%%%%%%%%%%%%%%%%%%%%%%%%%%%%%%%%%%%%%%%%%%%%%%%%%%%%

%%%%%%%%%%%%%%%%%%%%%%%%%%% fig 10 %%%%%%%%%%%%%%%%%%%%%%%
%%%%%%%%%%%%%%%%%%%%%%%%%%%%%%%%%%%%%%%%%%%%%%%%%%%%%%%%%%

%%%%%%%%%%%%%%%%%%%%%%%%%%% fig 11 %%%%%%%%%%%%%%%%%%%%%%%
%%%%%%%%%%%%%%%%%%%%%%%%%%%%%%%%%%%%%%%%%%%%%%%%%%%%%%%%%%

%%%%%%%%%%%%%%%%%%%%%%%%%%% fig 12 %%%%%%%%%%%%%%%%%%%%%%%
\begin{figure}[t]
%\vspace{-0.5cm}
\begin{center}
\includegraphics[width=0.48\textwidth]{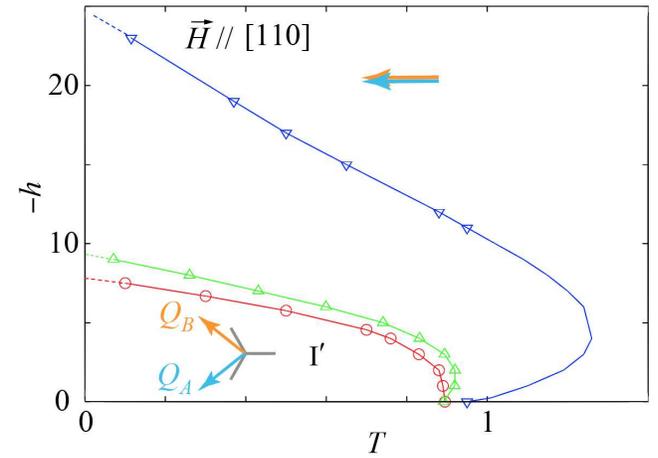}
\end{center}
%\vspace{-0.5cm}
\caption{
(Color online) 
$T$-$h$ phase diagram for $c$=0.1, 0.5, and 3.0  
under magnetic field ${\bf H} \parallel$ [110]. 
The dotted lines are the extrapolation to $T$=0. 
Quadrupole configuration is schematically shown 
by arrows along with the 3-fold axes.  }
\label{fig-HT110}
\end{figure}
%%%%%%%%%%%%%%%%%%%%%%%%%%%%%%%%%%%%%%%%%%%%%%%%%%%%%%%%%%

\subsection{$T$-$h$ phase diagram for $h<0$}\label{negativeh}
Let us now discuss the case $h<0$, {\it i.e.}, ${\bf H} \parallel [110]$. 
For $h<0$, the anisotropy due to the $c$ term does not compete with 
the conjugate field $h$ unlike the case of $h>0$. 
Thus, the unfavorable two domains in the phase I for $h>0$ 
become stabilized for $h<0$.
Figure \ref{fig-HT110} shows the $h<0$ part of 
$T$-$h$ phase diagram for $c=0.1,\ 0.5$, and $3.0$. 
The order-parameter configuration for one of the  
domains is schematically shown in Fig.~\ref{fig-HT110}.  
Since the canted configuration shown has energy gain from both of 
the anisotropy and the field, the two domains with $\langle Q_{u,+}\rangle<0$ are favored.
As for the anisotropy of the critical field, 
the zero-temperature critical field is $|h_c(0)|\sim 9$ for $c=0.5$, 
and this is about 50\% larger than the value for $h>0$ [see Fig.~\ref{fig-HT}(a)]. 
This is because the domains realized for $h<0$ are stabler against 
the applied field compared to the case of $h>0$.  
The anisotropy $c$ significantly enhances the 
reentrant behaviors as a function of $h$ in the phase diagram.  
The highest transition temperature is around 
$T_c\sim 1.25$ at $-h\sim 5$ for $c=3.0$, 
and this is about 40\% higher than the zero-field 
critical temperature $T_c (h=0)$.

\section{Discussion}
\label{sec-Discussion}

We have investigated phase transitions in the rotor 
and $\phi^4$ models with the $Z_3$ anisotropy 
and determined the $T$-$h$ phase diagram by MC simulations.  
In this section, we are going to discuss several points in more detail.  
We first carry out the Landau analysis for the phase boundaries in the 
$T$-$h$ phase diagram. 
Then, we will compare the present results to previous studies on 
the 3-state Potts model.\cite{Banavar,Grest,Rosengren,Kolesik,Pelizzola,Heilmann,Ni,Lapinskas,Rahman}
Finally, we will comment about recent experimental results 
in the Pr-based 1-2-20 compounds.

\subsection{Phase boundaries}
In Sec.~\ref{sec-THphase}, 
we have presented that the phase boundaries in the $h>0$ part have 
a different geometry from the result of the mean-field analysis.  
In the following, we will show that a simple phenomenological theory 
can describe most of the characteristic features in the $T$-$h$ phase 
diagram for both of the $h>0$ and $h<0$ cases.  

To describe the multiple ordered phases on the same footing, let us 
consider the instability of the para/polarized phase IV on the basis  
of the following Landau-type free energy, 
\begin{equation}
F = \sum_{s=A,B} V (\mathbf{Q}_s )  + \tilde{J} \mathbf{Q}_A  \cdot \mathbf{Q}_B, 
\end{equation}
where $V (\mathbf{Q}_s ) $ is the effective potential of the quadrupole 
moments on the sublattice $s$ and $\tilde{J} >0$ denotes the effective 
antiferro coupling between the two sublattice quadrupole moments.  
The effective potential is the generalization of the one considered 
in our previous work \cite{Hattori} by including the quadratic Zeeman 
coupling to applied magnetic field 
\begin{equation}
V( \mathbf{Q} ) 
= {\textstyle \frac{1}{2}} a \mathbf{Q}^2 
- {\textstyle \frac{1}{3}} \tilde{c} Q_u (Q_u^2 - 3 Q_v^2 ) 
+ {\textstyle \frac{1}{4}} \tilde{b} |\mathbf{Q}|^4 
- h Q_u. 
\end{equation}
Here, $\tilde{b}>0$ and $\tilde{c}>0$ are effective couplings 
and we assume that they are essentially constant in the part 
considered in the phase diagram.  

An essential difference from the previous study is that 
a finite moment is induced in the para/polarized phase IV, 
$ \langle \mathbf{Q}_s \rangle \equiv \bar{\mathbf{Q}} \equiv (\bar{Q},0)$, 
due to the conjugate field $h$, and symmetry breaking is 
related to the instability of fluctuations around this value, 
$\delta \mathbf{Q}_{A,B} \equiv \mathbf{Q}_{A,B}- \bar{\mathbf{Q}}$.  
The corresponding free energy is 
\begin{equation} 
\Delta F \sim \!\!\!
\sum_{s=A,B} \!\! {\textstyle \frac{1}{2}} 
( a_u \delta Q_{u,s}^2 + a_v \delta Q_{v,s}^2 ) 
+ \tilde{J} \delta \mathbf{Q}_A  \cdot \delta \mathbf{Q}_B + \cdots, 
\end{equation} 
where $\cdots$ is the higher order terms.  
The two components have different coefficients, 
when a finite moment is induced 
\begin{equation}
a_u = a - 2 \tilde{c} \bar{Q} + 3 \tilde{b} \bar{Q}^2 , \ \ 
a_v = a + 2 \tilde{c} \bar{Q} +  \tilde{b} \bar{Q}^2 . 
\label{eq:auav}
\end{equation}
Concerning the staggered mode $\delta \mathbf{Q}_- = 
\delta \mathbf{Q}_A - \delta \mathbf{Q}_B$, its 
$u$ component becomes unstable when $a_u \le \tilde{J}$, while 
$v$ component becomes unstable when $a_v \le \tilde{J}$.  
The former case corresponds to the ordered phase II, and 
the latter case corresponds to the ordered phase III.  
When both components are unstable, this is the phase I.  
Thus, the geometry of the phase boundaries is nothing but 
how the two lines, $a_u =\tilde{J}$ and $a_v =\tilde{J}$, 
position themselves.  

The result (\ref{eq:auav}) is very informative.  
It should be noted that the induced moment $\bar{Q}$ has the same 
sign as the conjugate field $h$ and its amplitude $| \bar{Q} |$ increases 
with $|h|$.  
The common term $a$ is the inverse of local quadrupole susceptibility 
at $h=0$, and its value decreases with lowering temperature.  
These mean that the two boundaries, $a_u =\tilde{J}$ and $a_v =\tilde{J}$, 
cross at the $h=0$ critical temperature, and their low-temperature 
sides are ordered phases.  

What happens about the two boundaries depends on the sign of $\bar{Q}$, 
{\it i.e.}, the sign of $h$, which is related to the magnetic field direction.  
In the case of $h<0$, one should notice that $a_v < a_u$ always holds.  
Therefore, the transition of the phase IV is always into the phase III. 
This is consistent with our result in Fig.~\ref{fig-HT}.  
It is not a problem that a sequential transition to the phase I 
does not occur.  
This is because the above result of $a_u$ is obtained with 
assuming $\bar{Q}_v=0$ and this no longer holds inside the phase III. 

The more exotic geometry of the phase boundaries for $h>0$ can be 
also explained by using a similar argument.  
As $h>0$ increases, the leading instability changes.  
$a_u < a_v$ for $0 \le \bar{Q} \le \bar{Q}_1$ 
and $a_v \le a_u$ for $\bar{Q} \ge \bar{Q}_1$, 
and the leading instability is the mode with the smaller $a$.  
Here, $\bar{Q}_1 = 2\tilde{c}/\tilde{b}$.  
This means that the two boundaries, 
$a_u =\tilde{J}$ and $a_v =\tilde{J}$, cross to each other twice 
(at $\bar{Q}$=0 and $\bar{Q}_1$) 
and the higher-field crossing is a  tetracritical point. 
With lowering temperature, the phase IV changes to the phase III 
at small $h$, while to the phase II at large $h$.  
This explains the exotic geometry of the phase boundaries in the 
$h>0$ phase diagram in Fig.~\ref{fig-HT}.  
It is also explained that the tetracritical point moves toward 
larger $h$ side for larger anisotropy $c$.  
These transitions are a $Z_2$ symmetry breaking except at the tetracritical point. 
Therefore, they should belong to the $3d$-Ising universality class.

Another interesting character in the phase diagrams is the reentrant 
behavior in the low-field part irrespective of the sign of $h$, and 
we can also explain this.  
The leading instability in the phase IV is determined by the 
smaller one of the two coefficients (\ref{eq:auav}). Note that  
 $\min \{a_u , a_v \} < a$ due to the linear 
term in $\bar{Q}$ as far as $\bar{Q}$ is not so large.  
Therefore, the local quadrupole susceptibility is enhanced by $h$, as far as 
$|h|$ is weak, and this means that the ordered phase expands with $|h|$.  
This explains the reentrant phase boundary in the low-field parts.  
This argument also predicts that the reentrant behavior is 
more prominent for $h<0$. 
This is because $a_v$ determines the phase 
boundary then and its reduction due to $\bar{Q}$-linear term 
persists up to a larger $|h|$ since the second-order increase $\bar{Q}^2$ 
has a smaller coefficient. 
Thus, this also agrees with our MC result.

\subsection{Comparison to the antiferro 3-state Potts model}
As we have discussed in Secs.~\ref{sec-zeromag} and \ref{sec-LT}, 
the present antiferro rotor model with $Z_3$ anisotropy shows a single 
phase transition at a finite temperature when $h=0$, 
and it belongs to the $3d$-XY universality class. 
The result is qualitatively the same also 
for the $\phi^4$ model with the third-order anisotropy discussed in Sec.~\ref{sec-THphase}. 
This indicates that the $Z_3$ anisotropy is irrelevant for the 
critical phenomena in these kinds of models.

It has been recognized that the anisotropy is not completely 
irrelevant.  
In the renormalization-group processes, 
the effects of the anisotropy become strongly suppressed 
near the critical fixed point but start to grow as the system approaches 
the low-temperature fixed point.\cite{Blankschtein,Oshikawa}
There, the macroscopic degeneracy inherent in the antiferro 3-state 
Potts model is lifted, and the $Z_3$ symmetry is broken 
together with the $Z_2$ sublattice symmetry.  
Remember that the ``3" in $Z_3$ corresponds to the three minima 
in the potential energy due to the 3-fold anisotropy.
This seems to happen unless the anisotropy $c$ is exactly zero. 
The symmetry broken state in our models is naturally connected 
to the PSS state proposed for the 3-state Potts 
model \cite{Rosengren,Lapinskas}. 
We expect that an effective coarse-grained theory for the Potts model 
on the diamond lattice is equivalent to the present one. 

Another interesting direction is to examine the same effective 
quadrupole models on different structures such as a simple cubic or 
body-centered cubic lattice, and this is related to a question 
if the ordered phase at $h=0$ has a universal character independent 
of details of bipartite lattice structures.  
Rahman {\it et al.}\cite{Rahman} claimed that the PSS state is not stable 
on a simple cubic lattice and the RS state is realized instead.  
For example, PrPb$_3$ is known as a quadrupole system 
with a simple cubic structure\cite{PrPb3} 
and it is interesting to study if there are qualitative differences 
between that case and the present one.
However, we should note that it is not appropriate to use 
our quadrupole models on a simple cubic lattice for 
discussing experimental results in PrPb$_3$.  
As emphasized in our previous work, intersite interactions 
are isotropic in the $Q_u$-$Q_v$ space only if the all bonds are  
along [111] or its equivalent directions under $T_d$ symmetry.  
This is not the case in a simple-cubic lattice, and 
therefore the intersite interactions are anisotropic 
and this anisotropy depends on the bond direction, 
which is similar to dipole-dipole interactions in this sense. 
One should use this type of quadrupole model with 
``dipole-type'' interactions to discuss quadrupole orders 
in PrPb$_3$. Indeed, these anisotropic interactions are known to be very 
 important in a context of the orbital orders in LaMnO$_3$\cite{Okamoto,Trebst}.

\subsection{Comparison to the Pr-based 1-2-20 systems}
Recently, various Pr 1-2-20 compounds have been found 
to have interesting low-temperature states.\cite{OniKusuReview} 
Here, we discuss their phase diagram under magnetic field 
based on our results in this paper.

First, let us discuss experiments on 
Pr{\it T}$_2$Zn$_{20}$, $(T$=Ir, Rh).
Although no experiment has directly identified its order parameter, 
it is believed an antiferro quadrupole order, since 
the results of the ultrasonic experiments \cite{Ishii1,Ishii2} show properties 
similar to those with the antiferro quadrupole order in other systems 
\cite{PrPb3}. 
Several experiments have reported the existence of multiple phases 
in magnetic field \cite{Onimaru0, Ishii1,Ishii2,Ikeura}. 
Our results predict their order parameters.  
As for the field direction dependence of  the phase diagram, 
our results qualitatively agree with the experimental ones, 
and also explain that critical temperatures are higher for ${\bf H}\parallel $ [110] 
than for ${\bf H} \parallel$ [001]. 
This holds universally as far as the sign of the $Z_3$ anisotropy 
is $c>0$. 
This indicates that the anisotropy $c$ is crucial 
for determining the field-direction dependence of the critical fields.

Another compound PrV$_2$Al$_{20}$\cite{Shimura,Tsujimoto,Shimura2} 
shows a double transition at $\mathbf{H}={\bf 0}$ \cite{Tsujimoto}.
Our results do not reproduce such a double transition at zero field, 
and this suggests that this system has other important interactions.
In the present models, the zero-field ordered state is extremely sensitive to the anisotropy
  and conjugate field $h$, and thus, tiny extrinsic strain or disorders mask the true $h=0$ properties, 
  which might realize as a double transition. For clarifying these issues, we need further efforts 
  and detailed analysis about such effects.

\section{Conclusion}
\label{sec-Summary}

In this work, we have analyzed two antiferro quadrupole models 
by using classical Monte Carlo simulations. 
The first model describes the order parameter with a two-dimensional 
degrees of freedom. 
It is reduced to the second model of an antiferro rotor type 
with a 3-fold anisotropy, 
and both of them are related to the antiferro 3-state Potts
model on the diamond lattice. 
We have clarified the universality class of the phase transition 
at zero field in these models is that of three-dimensional XY class,
which is consistent with previous studies for related models. 
One important point of our results is that 
the low-temperature ordered state is similar to the 
permutationally symmetric sublattice (PSS) state proposed 
for the antiferro 3-state Potts model. 
Another important result is that the scaling of the secondary 
order parameter is exotic and has a new set of critical exponents 
$\nu'$=0.597(12), $\eta'$=1.727(12), and $\beta'=\nu'(1+\eta')/2\simeq 0.815$.  
The secondary order parameter was first discovered in our 
previous study\cite{Hattori}.  
It is the ferro component of quadrupole and induced 
by the antiferro primary order parameter 
through the local $Z_3$ anisotropy.

\begin{figure}[t]
\begin{center}
\includegraphics[width=0.4\textwidth]{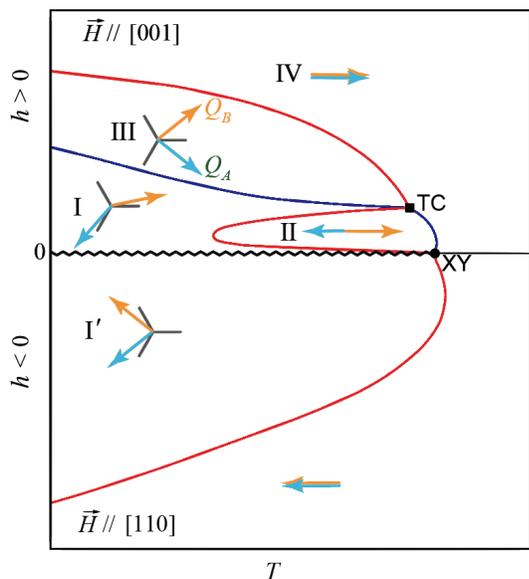}
\end{center}
\caption{
(Color online) 
Schematic $T$-$h$ phase diagram for an intermediate $c>0$. 
$h$ is related to magnetic field $\mathbf{H}$ as 
$h \propto 2H_z^2-H_x^2-H_y^2$, and the results for the two directions 
are combined. 
Quadrupole configurations are schematically 
shown by arrows with the 3-fold axes for the phase I, I$'$, and III. 
The zigzag line indicates first order transition at $h=0$ 
and it terminates at the critical end point XY, where 
the transition belongs to the $3d$-XY universality class.  
All the other lines correspond to the continuous transition 
of the $3d$-Ising universality class.  
The point TC is a tetracritical point.  
$R_v$ symmetry breaks on the red lines, while $R_{AB}$ 
symmetry breaks on the blue lines. 
}
\label{fig-schematicPhase}
\end{figure}

We have also investigated the effects of magnetic field.  
By taking into account the quadratic Zeeman coupling of 
quadrupoles to magnetic fields, we have determined the 
temperature-field phase diagram for two typical directions of magnetic field.  
A schematic picture of the result is plotted in 
Fig.~\ref{fig-schematicPhase} for an intermediate strength 
of the anisotropy, and 
we should note that the phase diagram is particularly rich 
for $\mathbf{H}$ $\parallel$ [001].   
With varying the anisotropy, we have further explored the 
phase diagram and found various multicritical points, 
which emerge as special points where multiple phases meet: 
bicritical and tetracritical points. 
Another important detail is reentrant behavior of the phase 
boundary near zero field, and this is prominent for larger anisotropy.  
We have succeeded in qualitative understanding of these phase structures and 
reentrant behaviors based on a simple phenomenology.   
Our results predict experimental realizations 
of multiple ordered states under
magnetic field and their multicriticality in interacting quadrupole systems.

%%%%%%%%%%%%%%%%%%%%%%%%%%% fig 13 %%%%%%%%%%%%%%%%%%%%%%%
%%%%%%%%%%%%%%%%%%%%%%%%%%%%%%%%%%%%%%%%%%%%%%%%%%%%%%%%%%

\section*{Acknowledgment}
The authors thank Tsuyoshi Okubo and Hiroshi Watanabe for 
fruitful discussions.  
This work is supported by a Grant-in-Aid for
Scientific Research [Grant No. 23740258 and No. 16H01079 (J-Physics)] 
from the Japan Society for the Promotion of Science.

\end{document}